\documentclass[]{resources/aa}

\usepackage{graphicx}
\usepackage[varg]{txfonts}
\usepackage{physics}
\usepackage{amsmath, amssymb}
\usepackage{dsfont}
\usepackage{bbm}
\usepackage{xcolor}
\usepackage{ulem}
\usepackage{xpatch}
\usepackage{hyperref}
\usepackage{import}
\usepackage{makeidx}
\usepackage{changes}
\usepackage{footnote}
\usepackage{placeins}

\usepackage{natbib}
\bibpunct{(}{)}{;}{a}{}{,} 

\newcommand{\sref}[1]{Sec.~\ref{#1}}
\newcommand{\tab}[1]{Table~\ref{#1}}
\newcommand{\fig}[1]{Fig.~\ref{#1}}
\newcommand{\equ}[1]{Eq.~(\ref{#1})}
\newcommand{\equo}[1]{Eq.~\ref{#1}}
\newcommand{\equs}[2]{Eqs.~(\ref{#1})~-~(\ref{#2})}
\newcommand{\Msunpyr}{\mathrm{M_\odot~yr^{-1}}}

\newcommand{\colout}[1]{\bgroup\markoverwith{\textcolor{#1}{\rule[.5ex]{2pt}{0.4pt}}}\ULon}

\makeatletter
\renewcommand*\aa@pageof{, page \thepage{} of \pageref*{LastPage}}
\newcommand{\pder}[2][]{\frac{\partial#1}{\partial#2}}
\makeatother

\makeindex

\begin{document}


%
\titlerunning{The influence of metallicity on a combined stellar and disk evolution}
\authorrunning{L. Gehrig et al.}
\title{The influence of metallicity on a combined stellar and disk evolution} 
\author{
 L.~Gehrig\inst{1},
 T.~Steindl\inst{2},
 E.~I.~Vorobyov\inst{1},
 R.~Guadarrama\inst{1},
 K.~Zwintz\inst{2}
}
\institute{
 Department of Astrophysics, University of Vienna,
 Türkenschanzstrasse 17, A-1180 Vienna, Austria
\and
Institut für Astro- und Teilchenphysik, Universität Innsbruck, Technikerstraße 25, A-6020 Innsbruck, Austria
}
\date{Received ....; accepted ....}
\abstract
{The effects of an accretion disk are crucial to understanding the evolution of young stars. During the combined evolution, stellar and disk parameters influence each other, motivating a combined stellar and disk model.
This makes a combined numerical model, evolving the disk alongside the star, the next logical step in the progress of studying  early stellar evolution.}
{We aim to understand the effects of metallicity on the accretion disk and the stellar spin evolution during the T~Tauri phase.}
{We combine the numerical treatment of a hydrodynamic disk with stellar evolution, including a stellar spin model, allowing a self-consistent calculation of the back-reactions between the individual components.}
{
We present the self-consistent theoretical evolution of T-Tauri stars coupled to a stellar disk. We find that
disks in low metallicity environments are heated differently and have shorter lifetimes, compared to their solar metallicity counterparts. Differences in stellar radii, the contraction rate of the stellar radius, and the shorter disk lifetimes result in faster rotation of low metallicity stars.}
{We present an additional explanation for the observed short disk lifetimes in low metallicity clusters. A combination of our model with previous studies (e.g., a metallicity-based photo-evaporation) could help to understand disk evolution and dispersal at different metallicities. Furthermore, the stellar spin evolution model includes several important effects, previously ignored (e.g., the stellar magnetic field strength and a realistic calculation of the disk lifetime) and we motivate to include our results as initial or input parameters for further spin evolution models, covering the stellar evolution towards and during the main sequence.}
\maketitle
%

%
%
%
%


\section{Introduction}
\label{sec:intro}


In the classical pre-main sequence (pre-MS) evolution, a star forms with a large radius and contracts isotropically along the Hayashi track \citep[][]{Hayashi61}.
During the contraction the star heats up due to the release of gravitational energy, nuclear reactions are triggered, a radiative core develops and the star evolves toward the main sequence (MS).
Within this standard model, however, the way the star gets its mass is not taken into account.

Within a collapsing cloud, two hydrostatic cores form \citep[e.g.,][]{Larson69, Bhandare18}. 
Starting from the second hydrostatic core (in the context of this work, the stellar seed), material from the parent cloud and the forming accretion disk accretes on the young star \citep[e.g.][]{Mercer-Smith84, Palla92, Hartmann97, Baraffe09, Dunham12,Kunitomo17, Vorobyov17c, Steindl21}.
The nature of the accretion process depends on different parameters (e.g., parent cloud mass, temperature, metallicity, or rotational rate), resulting in a variety of possible accretion rates over time \citep[e.g.,][]{vorobyov10, Vorobyov17c, Vorobyov20}.

The pre-MS evolution starting from the stellar seed is strongly influenced by the amount of energy added to the stellar interior ($L_\mathrm{add}$) during the accretion process \citep[e.g.,][]{Kunitomo17, Steindl21}.
If the accreting material loses its energy during the accretion process ($L_\mathrm{add}\rightarrow 0$, cold accretion), the stellar radius is significantly (up to one order of magnitude) smaller compared to the classical pre-MS evolution \citep[e.g.,][]{Kunitomo17}.
If, on the other hand, the accreting material can add its energy in the stellar interior ($L_\mathrm{add}$ large, hot accretion), the stellar radius inflates and can become comparable with classical evolutionary models.
After accretion has finished and the energy added into the stellar interior ceases, the star can evolve along the Hayashi track \citep[e.g.,][]{Kunitomo17}.
We note that the exact pre-MS evolutionary track, starting from a stellar seed, depends on how much energy can be added to the star and where this energy is added in the stellar interior \citep[e.g.,][]{Kunitomo17, Steindl21}.
Besides the cold and warm accretion models, $L_\mathrm{add}$ can also depend on the stellar accretion rate \citep[hybrid accretion, e.g.,][]{Vorobyov17c}.
In periods of high accretion rates, $L_\mathrm{add}$ is large (similar to hot accretion) and during low accretion rates $L_\mathrm{add}$ stays small (similar to cold accretion).

During the early (Class~0 and Class~I) phases of stellar evolution, the accretion rate on the star can vary significantly over time. 
During episodic outburst events (e.g., FU~Orionis outbursts), the stellar accretion rate can rise over orders of magnitude during a short timescale \citep[$\sim$~years, e.g.,][]{vorobyov10,Vorobyov20}.
With $L_\mathrm{add}$ coupled to the accretion rate, episodic outburst events during the early disk phase (e.g., FU~Orionis outbursts) can lead to radial oscillations \citep[e.g.,][]{Bastien11, Vorobyov17c}, affecting the stellar pre-MS evolution.
In previous studies, the accretion history on the star is either assumed to be constant over time or pre-computed \citep[e.g.,][]{Kunitomo17, Vorobyov17c, Jensen2018, Elbakyan2019, Steindl21}.
In the latter case, the effect of episodic outbursts can be included.
Up to now, however, such calculations only deal with the stellar structure and are not able to include any back-reaction of the evolving star on the accretion disk.

Star-disk interactions furthermore influence the stellar rotation period.
Understanding the stellar spin evolution allows the determination of stellar ages \citep[gyrochronology,][]{Barnes07}, magnetic activity or high-energy stellar radiation \citep[e.g.,][]{Pallavicini81, Micela85, Pizzolato03, Wright11, France18}. 
During the disk phase, the stellar rotation period can also influence the evolution of the accretion rate. 
Fast-rotating stars tend to experience more outbursts compared to slow-rotating stars \citep[e.g.,][]{Gehrig2022}.
According to recent spin evolution models \citep[e.g.,][]{Matt10, Matt12, Gallet19, Ireland21}, the stellar accretion rate is an important factor to determine the stellar rotation rate. A precise calculation of the accretion rate is crucial for understanding the rotational evolution of young stars.
Furthermore, stellar metallicity has an effect on the stellar spin evolution \citep[e.g.,][]{Amard19, Amard20}. 
In their work, \cite{Amard19} present a grid of low-mass ($\leq 1.5~\mathrm{M_\odot}$) isochrones including a spin evolution model.
The disk phase, however, is treated in a simplified way.
During the disk phase, the stellar period is assumed to be constant (disk-locking) and the disk lifetime is a free parameter and independent of stellar mass and metallicity.
This assumption is in conflict with observations of disk fractions in low metallicity clusters \citep[e.g.,][]{Yasui10,Yasui16,Yasui21,Guarcello21} and theoretical, metallicity-dependent photo-evaporation models \citep[e.g.,][]{Nakatani18}, indicating short disk lifetimes in low metallicity environments.

In this study, we aim to combine "The Adaptive Implicit RHD" (TAPIR) code \citep[][]{ragossnig20, Steiner21} with the software instrument Modules for experiments in stellar astrophysics \citep[MESA, ][]{paxton2011, paxton2013,paxton2015,paxton2018, paxton2019} to provide a self-consistent numerical description of a star-disk system at a young age.
This allows us to study, which parameters affect, e.g., the disk lifetime, the stellar spin, and the pre-MS evolution, including the back-reaction on the other component.
In this first paper, introducing the new method, we want to study the effects of metallicity on the disk and stellar spin evolution of T~Tauri stars (with ages $\gtrsim 1$~Myr). 
We seek to understand how stellar metallicity affects the accretion disk and stellar spin evolution of T~Tauri stars.

This paper is structured as follows: \sref{sec:model_description} describes the disk and stellar evolution models used in this study. Moreover, the combination of the disk, spin, and stellar model is introduced.
Our results are presented in \sref{sec:results} and discussed in \sref{sec:discussion}. Finally, we draw our conclusions in \sref{sec:conclusion}.

%
%
%
%
%
%


\section{Model description}
\label{sec:model_description}

\subsection{Hydrodynamic disk evolution with the TAPIR code}\label{sec:hydrodynamic_disk_evolution}

Studying the long-term evolution of an accretion disk including the inner disk regions close to the star, we use the implicit TAPIR code, which is described in detail in \cite{ragossnig20} and \cite{Steiner21}. The key advantage of our numerical method is the treatment of hydrodynamic equations, rather than a diffusive disk evolution approach \citep[e.g.][]{pringle81,Armitage01, zhu07, Schib21}. We can include the influence of pressure gradients and a stellar magnetic field, which is of particular importance in the inner disk regions \citep[e.g.][]{Romanova02, bessolaz08}. The following summarizes some key features of the TAPIR code.

\textit{Basic equations:} 
In our model, we use a time-dependent, vertically integrated, viscous accretion disk formulation \citep[e.g.][]{Shakura1973, Armitage01, Steiner21}.
\begin{alignat}{2}
    & \pder{t} \, \Sigma &&+ \nabla \cdot ( \Sigma \, \vec u ) = 0\;, \label{eq:cont} \\
    & \pder{t} (\Sigma \, \vec u) &&+ \nabla \cdot (\Sigma \, \vec u : \vec u) - \frac{B_\mathrm{z} \vec B}{2 \pi}\nonumber \\
    & &&+ \nabla P_\mathrm{gas} + \nabla \cdot Q + \Sigma \, \nabla \psi + H_\mathrm{P} \nabla \left( \frac{B_\mathrm{z}^2}{4 \pi} \right) = 0 \;, \label{eq:mot} \\
    &\pder{t} (\Sigma \, e) &&+ \nabla \cdot (\Sigma \, \vec u \, e ) + P_\mathrm{gas} \, \nabla \cdot  \vec u \nonumber \\
    & &&+ Q : \nabla \vec u - 4 \pi \, \Sigma \, \kappa_\mathrm{R}\left(J - S \right) + \dot E_\mathrm{rad} = 0 \;, \label{eq:ene}
\end{alignat}
where $\Sigma$, $\vec u$, $e$, $P_\mathrm{gas}$ and $H_\mathrm{P}$ stand for the gas column density, gas velocity in the planar components $\vec u = (u_\mathrm{r}, u_\mathrm{\varphi})$, specific internal energy density per surface area, vertically integrated gas pressure and the vertical scale height of the gas disk, respectively.
The gradient in planar cylindrical coordinates reads $\nabla = (\partial / \partial r , r^{-1} \partial / \partial \varphi)$ with $\partial / \partial \varphi = 0$ for our axisymmetric model.
$P_\mathrm{gas}$ is utilized by the ideal equation of state $P_\mathrm{gas} = \Sigma e ( \gamma - 1 )$, with the adiabatic coefficient $\gamma = 5 / 3$ and $\kappa_\mathrm{R}$ denotes the Rosseland-mean opacity, which is composed of a gaseous component $\kappa_\mathrm{R,gas}$,
which are based on \cite{Ferguson2005} with the abundances in \cite{caffau11} \citep[see Fig.~1 in][]{Steiner21}, and a dust-dominated component $\kappa_\mathrm{R,dust}$ \citep[based on][]{pollack85}; with $\kappa_\mathrm{R} = \kappa_\mathrm{R,gas} + f_\mathrm{dust} \kappa_\mathrm{R,dust}$, where $f_\mathrm{dust}$ is the dust-to-gas mass ratio set to the respective stellar metallicity.
In our model, $f_\mathrm{dust}$ equals the stellar metallicity, ranging from $Z = 0.0028$ to $Z = 0.014$.
The gravitational potential of the star-disk system is denoted by $\psi$.
The vertical component of the stellar magnetic field $B_\mathrm{z}$ is assumed to be a dipole and constant within the disk's vertical extent. The planar magnetic field components $\vec B = (B_\mathrm{r}, B_\mathrm{\varphi})$ are taken at the surface of the disk. 
Similar to \cite{rappaport04} and \cite{kluzniak07}, we ignore the radial magnetic field component $B_{\rm r} = 0$. 
Differential rotation between the stellar magnetic field lines, which are assumed to be in rigid rotation with the star, and the disk's material generates an angular magnetic field component $B_{\rm \varphi}$ \citep[e.g.][]{rappaport04,kluzniak07,Steiner21,Gehrig2022}.
The radial radiative transport (depicted by $(J - S)$) is modeled in a radiative diffusion approximation with an Eddington factor of $1/3$.
We note that an Eddington factor of 1/3 is usually valid within optically thick regions or in optically thin regions in the presence of a uniform radiation field. We assume that the stellar radiation and the ambient temperature roughly meet this requirement. The method, however, has its weaknesses when transitioning between optically thick and thin regions and a variable method would produce better results. 
Here, $J$ and $S$ stand for the 0th moment of the radiation field and the source function, respectively \citep[e.g.][]{ragossnig20, Steiner21}.
Finally, $\dot E_\mathrm{rad}$ and $Q$ depict the net heating/cooling rate per unit surface area and the viscous stress tensor, respectively \citep[e.g.][]{Steiner21, Gehrig2022}. 
The kinematic viscosity is formulated according to \cite{Shakura1973} $\nu = \alpha c_\mathrm{S} H_\mathrm{P}$; with the viscous $\alpha$ parameters, the isothermal sound speed $c_\mathrm{S}$ and the pressure scale height $H_\mathrm{P}$.
In addition to \equs{eq:cont}{eq:ene}, we utilize an adaptive grid \citep[][]{Dorfi1987} to allow a moving inner and outer disk boundary as well as adequate radial gridpoint resolution throughout the computational domain.

\textit{Computational domain:}
The innermost disk regions are assumed to be truncated and eventually disrupted by a strong ($\sim$~kG) stellar magnetic field inside the so-called truncation radius $r_{\rm trunc}$ \citep[e.g.][]{Romanova02, bessolaz08, Romanova14}. With our one-dimensional code, the physical processes inside $r_{\rm trunc}$ can not be described. Thus, we choose $r_{\rm trunc}$ to be the inner boundary of our computational domain.
The radial position of $r_{\rm trunc}$ can be calculated by balancing the stellar magnetic pressure with the maximum from the ram pressure of the accreting disk material and the gas pressure \citep[e.g.][]{Koldoba02, Romanova02, bessolaz08, Steiner21, Gehrig2022}
\begin{equation}\label{eq:r_trunc}
    P_\mathrm{magn}(r_\mathrm{trunc}) = \text{max}\left[P_\mathrm{ram}(r_\mathrm{trunc}), P_\mathrm{gas}(r_\mathrm{trunc})\right]\;.
\end{equation}
In case $P_\mathrm{ram} \geq P_\mathrm{gas}$, the truncation radius can be calculated following \cite{hartmann16}
\begin{align}
    r_{\mathrm{trunc}}(P_\mathrm{ram} \geq P_\mathrm{gas}) \approx 18 \, \xi \, R_\odot & \, \left(\frac{B_\star}{10^3 \, G}\right)^{4/7}  \left(\frac{R_\star}{2 \, R_\odot}\right)^{12/7} \left(\frac{M_\star}{0.5 \, M_\odot}\right)^{-1/7} \nonumber \\ 
    & \left(\frac{\dot M_\star}{10^{-8} \, M_\odot / \mathrm{yr}}\right)^{-2/7} \label{eq:truncation_radius} \;, 
\end{align}
with the correction factor $\xi = 0.7$, the stellar dipole magnetic field strength $B_\star$, the stellar radius $R_\star$, the stellar mass $M_\star$ and the accretion rate on the star $\Dot{M}_\star$. 
In the case of $P_\mathrm{ram} < P_\mathrm{gas}$, the position of the truncation radius is calculated by equating the magnetic pressure of the vertical field component and the gas pressure
\begin{align}
    &r_{\mathrm{trunc}}(P_\mathrm{ram} < P_\mathrm{gas}) = \frac{{B_\star}^{1/3} \, R_\star}{{P_\mathrm{gas}}^{1/6} \, (8 \pi )^{1/6} } \;. 
    \label{eq:truncation_radius_press}
\end{align}
$r_{\rm trunc}$ is allowed to move during our simulations according to stellar and disk evolution. Furthermore, the outer boundary of our domain is set to the radial position where the surface density has reached a certain value $\Sigma_{\rm out} = 1.0~{\rm g/cm^2}$ \citep[similar to, e.g.,][]{Vorobyov20}.

\textit{Boundary conditions:} 
At the inner boundary, we choose "free"/Neumann boundary conditions \citep[e.g. $\partial \Sigma / \partial r = 0$,][]{Romanova02, Romanova04, Steiner21}. At the outer boundary, we fix the radial (angular velocity) $u_{\rm r}$ ($u_{\rm \varphi}$) to zero (the Keplerian value), respectively. This indicates no inflow of additional material from an outer envelope or parent cloud, as expected in late Class~II star-disk systems. The surface density $\Sigma$ and the specific internal energy $e$ are again treated with a Neumann boundary condition.

\subsection{Stellar evolution with MESA}\label{sec:MESA}

The stellar evolution calculations in this work are performed with version v12778 of \textit{Modules for Experiments in Stellar Astrophysics} \citep[MESA][]{paxton2011, paxton2013, paxton2015, paxton2018, paxton2019}. MESA is an open-source software instrument that solves the fully coupled structure and composition equations simultaneously for a spherically symmetric and hence one-dimensional stellar model \citep{paxton2011}. Details about the adopted micro-physics are given in Appendix~\ref{app:mesaphysics}.
The treatment of rotation in MESA is summarized in Appendix~\ref{app:mesarot}.

The input physics for the stellar evolution calculations performed in this work is identical to \citet{Steindl21}. The initial models used in this study are identical to the ones used in \citet{Steindl21} with a composition of $X_{^1\mathrm{H}} = 1 - X_{^2\mathrm{H}} - X_{^3\mathrm{He}} - X_{^4\mathrm{He}} - Z$, where $X_{^2\mathrm{H}} = 20$~ppm, $X_{^3\mathrm{He}} = 85$~ppm, $X_{^4\mathrm{He}} = 0.276$ and $Z = 0.014$ for the models with solar metallicity or $Z = 0.0028$ for models with a fifth of solar metallicity. In the following, we briefly discuss implemented changes from the usual MESA calculations for the treatment of accretion. 

We follow the description of \citet{Baraffe09} for which the accretion onto the protostar proceeds non-spherically, allowing the central object to freely radiate energy across most of the photosphere \citep{Hartmann97}. The energy budget of the accretion process is given by
\begin{equation}
    \frac{\mathrm{d} E_{\rm acc}}{\mathrm{d} t} = (\epsilon -1)\frac{G M_\star \dot M_\star}{R_\star}\,,
\end{equation}
which is a combination of the gravitational ($-G M_\star/R_\star$) and internal energy ($+\epsilon G M_\star/R_\star$) of the accreted material. The geometry of the accretion process is described by the free parameter $\epsilon$. We follow the approach of previous works \citep{Baraffe09, Jensen2018, Steindl21} and set $\epsilon = 0.5$, corresponding to accretion from a thin disc around the equator \citep{Hartmann97, Baraffe09}.
Thus, the total energy rate accreted to the star from a thin disk is given as $\epsilon G M_\star \dot{M_\star}/R_\star$.

The energy of the accreted material is either added to the star as extra heat ($L_{\rm add}$) or radiated away as accretion luminosity ($L_{\rm acc}$). 
The accretion luminosity is added to the intrinsic stellar luminosity and the additional heating of the disk is considered in \equ{eq:ene}.
A free parameter controls the corresponding amounts such that 
\begin{equation}
    L_{\rm add} = \beta\epsilon \frac{G M_\star \dot M_\star }{R_\star}
\end{equation}
and
\begin{equation}
    L_{\rm acc} = (1-\beta)\epsilon \frac{G M_\star \dot M_\star}{R_\star}.
\end{equation}
The value $\beta$ and its dependence on star and disk properties are free parameters. From earlier studies it is evident that $\beta$ varies with accretion rate \citep[see e.g.][]{Baraffe2012, Jensen2018, Elbakyan2019}. In this work, we use a functional dependency first presented by \citet{Jensen2018} given as
\begin{equation}\label{eq:beta_mesa}
    \beta(\Dot{M_\star}) = \frac{\beta_{1} \exp\left( \frac{\Dot{M}_{\rm m}}{\Delta}\right) + \beta_{2 } \exp\left( \frac{\Dot{M}_\star}{\Delta}\right)}{ \exp\left( \frac{\Dot{M}_{\rm m}}{\Delta}\right) + \exp\left( \frac{\Dot{M}_\star}{\Delta}\right)}
\end{equation}
This describes a step function with a smooth transition between $\beta_\mathrm{1}$ and $\beta_\mathrm{2}$. The functional dependency is controlled by four parameters: $\beta_{1}= 0.005$, $\beta_{2}= 0.2$, a midpoint accretion rate of $\Dot{M}_{\rm m} = 6.2 \cdot 10^{-6}~\Msunpyr$  and a crossover width of $\Delta= 5.95 \cdot 10^{-6}~\Msunpyr$. For comparability, we use the same values as in \citet{Steindl2022}, which were originally adapted from \citet{Jensen2018} for numerical stability. In our case, this translates to a constant value $\beta$ at the beginning of the evolution which becomes gradually smaller as the accretion rate decreases (see Figure \ref{fig:star_h_accr}).
During the calculation of the initial models, the accretion rates and thus $\beta(\Dot{M_\star})$ vary significantly.
Given the proportionately small accretion rates, $\beta$ is basically constant during the joint evolution of MESA and TAPIR.

The dependence of $\beta$ on the accretion rate can also be physically interpreted. 
During phases of low accretion rates, the stellar magnetic field can truncate the disk at several stellar radii, and the disk material is funneled on the star \citep[e.g.,][]{bessolaz08, hartmann16}.
The in-falling disk material generates a shock close to the stellar surface and most of the energy is radiated away \citep[e.g,][]{Koenigl91}.
As a consequence, only a small part of the disk material's energy is assumed to be added to the stellar interior.
On the other hand, during phases of high accretion rates, the disk is pushed towards the star, and accretion along magnetic field lines ceases.
Without the shocks at the end of the magnetic funnels, a larger amount of energy can be added to the star.
The transition between the regimes of low and high $\beta$ values is expected to occur at $\sim 10^{-6}~-~10^{-5}~\Msunpyr$ \citep[e.g,][]{Vorobyov17c, Jensen2018}, which motivates the choice of $\Dot{M}_{\rm m}$. 
We note, however, that the description of $\beta$ in \equ{eq:beta_mesa} does not take metallicity into account.
We assume that, in case of low accretion rates, the inner part of the disk is sufficiently ionized due to stellar irradiation.
Stellar magnetic field lines can couple to the disk material and the accretion geometry is unaffected by metallicity.
During phases of high accretion rates, when the disk is pushed towards the stellar surface, differences in disk ionization due to metallicity are assumed to be negligible as well.
The transition between the two accretion regimes, however, could be affected by the disk's ionization fraction, which depends on metallicity.
Thus, the value $\Delta$ should include a metallicity-dependent factor, that would exceed the frame of this study.

The heat added to the stellar model is distributed according to 
\begin{equation}
    l_{\rm extra} = \frac{L_{\rm add}}{M_\star} {\rm max} \left\{0, \frac{2}{M_{\rm outer}^2} \left(\frac{m_r}{M_\star} -1 + M_{\rm outer} \right)\right\}.
\end{equation}
which follows the approach of \citet{Kunitomo17}. In this picture, the heat is distributed only in an outer region of fractional mass $M_{\rm outer}$ and increases linearly with the mass coordinate $m_r$. In our study, we choose $M_{\rm outer} = 0.01$, hence injecting the heat only in the outer 1\% of the stellar structure.

\subsection{Stellar spin model}\label{sec:stellar_spin_model}

The stellar spin model used in our simulations is based on \cite{Matt10} and \cite{Gallet19} and summarized in detail in \cite{Gehrig2022}. 
In this study, we assume solid body rotation for the star.
Angular momentum transport within the star is discussed in \sref{sec:AM_within}.
The stellar angular momentum $J_\star = I_\star \Omega_\star$ is influenced by external torques $\Gamma_{\rm ext}$ and its temporal derivative reads
\begin{equation}\label{eq:stellar_angular_momentum}
    \Dot{J}_\star = I_\star \Dot{\Omega}_\star + \Dot{I}_\star \Omega_\star = \Gamma_\mathrm{ext} \, ,
\end{equation}
with the stellar angular momentum $\Omega_\star$ and moment of inertia $I_\star$.
Rearranging \equ{eq:stellar_angular_momentum} results in a time-dependent equation for the stellar spin
\begin{equation}\label{eq:stellen_spin_evolution}
    \Dot{\Omega}_\star = \frac{\Gamma_\mathrm{ext}}{I_\star} - \frac{\Dot{I}_\star}{I_\star} \Omega_\star \, .
\end{equation}
The stellar moment of inertia $I_\star$ is calculated by the MESA code.
Within $\Gamma_{\rm ext}$ the effects of accretion $\Gamma_\mathrm{acc}$, stellar winds $\Gamma_\mathrm{W}$ \citep[in form of an accretion powered stellar wind APSW,][]{Matt05b} and the magnetic star-disk connection $\Gamma_\mathrm{ME}$ \citep[in form of magnetospheric ejections MEs,][]{Zanni13} are combined. 
During the process of accretion, the star gains mass and angular momentum from the disk that increases $I_\star$ and spins up the star, respectively. 
An APSW, on the other hand, is assumed to eject a specific amount of the accreted material ($\Dot{M}_{\rm W} = W \Dot{M}$) removing stellar mass and angular momentum. 
$W$ is expected to be $\lesssim 2$~\% \citep[e.g.][]{Crammer08, Pantolmos20} and thus, we choose $W=2$~\% in this study.
The effect of MEs on the stellar spin evolution depends on the position of $r_{\rm trunc}$ with respect to the co-rotation radius
\begin{equation}
    r_\mathrm{cor} = \left( \frac{G \, M_\star}{\Omega_\star^2} \right)^{1/3} \label{eq:corotation_radius} \;,
\end{equation}
with the gravitational constant $G$ and the stellar mass $M$. 
MEs increase the stellar angular momentum and spin up the star if $r_\mathrm{trunc} < K_\mathrm{rot}^{2/3} r_\mathrm{cor}$. Otherwise, the star spins down. Following \cite{Gallet19}, we use $K_\mathrm{rot} = 0.7$ in this study\footnote{We note that in more recent studies of MEs \citep[e.g.][]{Pantolmos20,Ireland21} smaller values for $K_{\rm rot}$ are found. $K_{\rm rot}=0.7$ can be understood as an upper limit \citep[][]{Ireland21}.} and the external torque contributions scale as:
\begin{equation}
    \Gamma_\mathrm{acc} = K_\mathrm{acc} \Dot{M}_\mathrm{\star} r_\mathrm{trunc}^2 \Omega_\mathrm{disk,in} \, 
\end{equation}
\begin{equation}\label{eq:gamma_wind}
    \Gamma_\mathrm{W} = \Dot{M}_\mathrm{W} r_\mathrm{A}^2 \Omega_\star  \,\, \mathrm{and}
\end{equation}
\begin{equation}\label{eq:gamma_me}
    \Gamma_\mathrm{ME} = K_\mathrm{ME} \frac{B_\star^2 R_\star^6}{r_\mathrm{trunc}^3} \left[ K_\mathrm{rot} - \left( \frac{r_\mathrm{trunc}}{r_\mathrm{cor}} \right)^{3/2} \right] \, ,
\end{equation}
with $K_\mathrm{acc}= 0.4$, $K_\mathrm{ME}=0.21$, the disk rotation rate at the truncation radius $\Omega_\mathrm{disk,in}$, and the stellar Alfv\'en radius 
\begin{equation}\label{eq:alfven}
    r_\mathrm{A} = K_\mathrm{1} \left[ \frac{B_\star^2 R_\star^2}{\Dot{M}_\mathrm{wind} \sqrt{K_\mathrm{2}^2 v_\mathrm{esc}^2 + \Omega_\star^2 R_\star^2} }  \right]^m R_\star \, ,
\end{equation}
where $v_\mathrm{esc} = \sqrt{2 \mathrm{G} M_\star / R_\star}$, $m = 0.2177$, $K_\mathrm{2} = 0.0506$, and $K_\mathrm{1}~=~1.7$.

\subsection{Numerical combination of the disk, spin and stellar model}\label{sec:MESA_TAPIR_combination}

\begin{figure}
    \centering
    \resizebox{\hsize}{!}{\includegraphics[trim={4.3cm 2.4cm 7.6cm 7.4cm},clip]{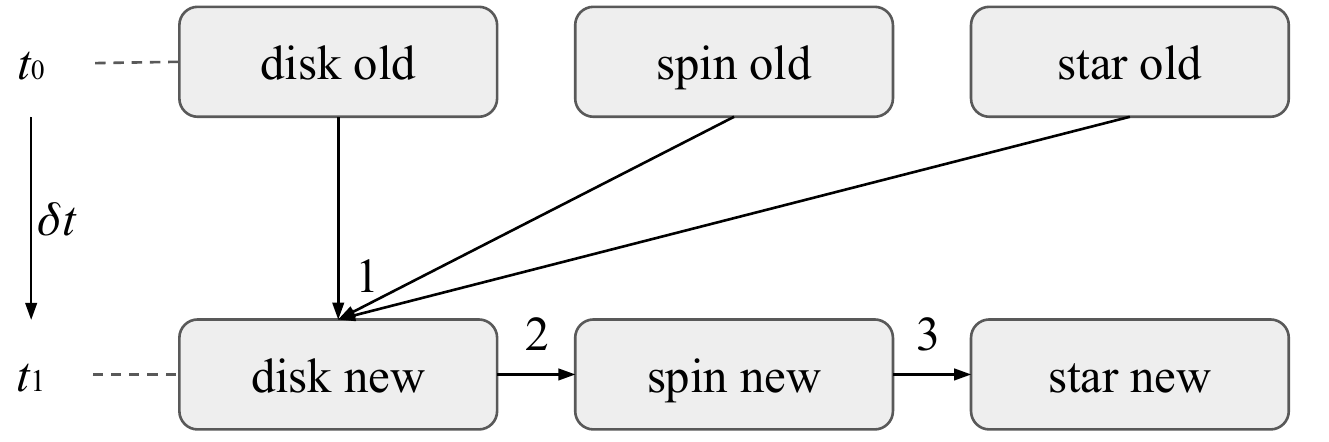}}
    \caption{Schematic representation of the interaction between TAPIR and MESA. Starting from an initial/old timestep $t_0$, the disk, spin and star values are evolved towards the new timestep $t_1 = t_0 + \delta t$. The new values at $t_1$ are obtained in three steps, represented by the numbered arrows (see text).}
    \label{fig:sketch}
\end{figure}

The interaction between the disk (TAPIR, \sref{sec:hydrodynamic_disk_evolution}), spin (stellar spin model, \sref{sec:stellar_spin_model}), and stellar model (MESA, \sref{sec:MESA}) is schematically shown in \fig{fig:sketch}. 
Starting from the initial or old values at timestep $t_\mathrm{0}$, the new disk values, spin values, and stellar values at timestep $t_\mathrm{1} = t_\mathrm{0} + \delta t$ are calculated in 3 steps, which are represented by numbered arrows in \fig{fig:sketch}. 
We note that we calculate the stellar spin evolution and the MESA evolution separately.
During the MESA calculation, the star is assumed to rotate at a constant rate.
In the first step, the TAPIR code calculates the disk evolution from timestep $t_\mathrm{0}$ to $t_\mathrm{1}$ using the old disk, spin, and stellar values (Arrow 1). 
Then, the spin values are updated based on the new disk values (Arrow 2).
Finally, the MESA code is updated and evolves the stellar parameters from timestep $t_\mathrm{0}$ to $t_\mathrm{1}$ (Arrow 3).

The accuracy and reliability of our method depend on the timestep size $\delta t$. 
Choosing small timesteps results in accurate but long computational times.
Large timesteps will reduce computational time but accuracy suffers.
We choose $\delta t = 100$~yr in this study. 
During 1~Myr the stellar values are updated $10^4$ times resulting in an acceptable representation of the stellar evolution. 
We also tested a range of $\delta t$ from 1~yr to 1000~yr and found no significant change in our results.
We note that MESA and TAPIR use multiple (internal) timesteps to evolve the star and the disk from time $t_\mathrm{0}$ to $t_\mathrm{1}$.
For MESA, the internal timesteps range between $\sim 10^{6}$~s and $\sim 10^{8}$~s. 
For TAPIR, the timesteps range between $\sim 10^4$~s and $\sim 10^8$~s \citep[see Fig.~A.1 in][]{Steiner21}.
We note that we do not model for example episodic outburst events in this study. 
During these outbursts, the accretion rate onto the star can change significantly within short time scales and influence the stellar evolution \citep[e.g.][]{Vorobyov17c}. 
In further studies, such outbursts are included in our model and $\delta t$ has to be adapted to lower values to assure a sufficiently accurate stellar evolution calculation.

%
%
%
%
%
%


\section{Results}
\label{sec:results}

To study the influence of metallicity on the combined evolution of a T~Tauri star and its disk, several stellar and disk parameters have to be discussed.
These parameters are defined based on observations, and previous theoretical or numerical studies and, within a reasonable range, initial models for a T~Tauri star and its disk are constructed.
We choose $t_{\rm 0} = 2$~Myr as an initial time for our combined simulations.
The effects of the early evolution and different values for $t_{\rm 0}$ are discussed in \sref{sec:early_evo}.
For each model, the stellar metallicity is varied between $Z = 0.2~\mathrm{Z_\odot}$ \citep[motivated by observations of young, low metallicity clusters][]{Yasui10, Yasui16, Yasui21} and $Z = 1~\mathrm{Z_\odot}$.

\textit{Disk parameters}: One important parameter that defines an accretion disk is the mass accretion rate. For T~Tauri star-disk systems ($\lesssim 10$~Myr) accretion rates range within $10^{-10}$ and $10^{-7}~\Msunpyr$ \citep[e.g.][]{Gullbring98, Manara16, Vorobyov17c, Manara22}. 
For  $t_{\rm 0} = 2$~Myr, \cite{Manara2012} and \cite{Testi2022} find disk accretion rates for a $\sim 1~\mathrm{M_\odot}$ star between $10^{-8}~\Msunpyr$ and $10^{-9}~\Msunpyr$.
Furthermore, the viscous $\alpha$-parameter can have a wide range between $0.1$ and $\sim 0.0001$ \citep[e.g.][]{zhu07, zhu09b, Vorobyov09, Mulders2012, Pinte2016, Yang18, Flaherty2020}. 
In our simulations, we choose the initial (starting) accretion rate $\Dot{M}_{\rm start} = [1\times10^{-9},5\times10^{-9},1\times10^{-8}]~\Msunpyr$ and set the (constant) viscous $\alpha$-value to 0.01 unless otherwise stated.
This combination results in the expected disk lifetimes of $\lesssim 10$~Myr \citep[consistent with e.g.][]{Richert2018}. 
We stop our simulations when the disk accretion rate has decreased to $10^{-11}~\Msunpyr$. 
This lower limit is motivated by the lowest detectable accretion rate for a $1~\mathrm{M_\odot}$ star \citep[e.g.,][]{Sicilia16}. 
At such low accretion rates, the disk is expected to be dissolved rather quickly due to photo-evaporation, which is currently not included in our model.
We note that an adaptive viscosity description will be added in further studies.

\textit{Stellar parameters:} The inner disk region close to the star is influenced by a stellar magnetic field. The field strength of young T~Tauri stars can vary between several hundred to several thousand Gauss \citep[e.g.][]{Johnstone14, Lavail17, Lavail19}. Unless otherwise stated, the magnetic field strength is chosen to be 2.0~kG. 
In addition, stellar parameters such as radius, luminosity and effective temperature at $t_\mathrm{0}$ result from the initial stellar evolution ($t<t_\mathrm{0}$).
The initial stellar parameters are chosen as described in \cite{Kunitomo17}, which follow \cite{Stahler1988} and \cite{Hosokawa2011}.
We start from a fully convective stellar seed with a mass of $0.01~\mathrm{M_\odot}$ ($\approx$~10 Jupiter masses) and a radius of $1.5~\mathrm{R_\odot}$ \citep[e.g.][]{Kunitomo17, Steindl21}. 
As pointed out in \cite{Hosokawa2011}, the initial entropy of the stellar seed influence the subsequent stellar evolution in the case of cold accretion. 
Cold accretion, however, is assumed to be unrealistic for most stars \citep[e.g.,][]{Kunitomo17, Vorobyov17c}.
In the case of hot or warm accretion, the differences due to a variation in the initial stellar seed parameters are expected to be small \citep[e.g.,][]{Stahler1988, Kunitomo17}.
Thus, we do not include a variation of the initial stellar seed parameter in this study and refer to Appendix~D in \cite{Kunitomo17} for further information.
Each model accretes material until it has reached a stellar mass of $0.91$~$\mathrm{M_\odot}$ and the desired accretion rate $\Dot{M}_\mathrm{start}$ at time $t_{\rm 0} = 2$~Myr. The effect of the accretion history is described in \sref{sec:star_init}.
During the T~Tauri phase, the rotational period can span over an order of magnitude from $\lesssim 1$ towards $\sim 10$~days \citep[e.g.][]{Serna2021}. 
To compare the effect of different metallicity values, we start all our models at the same initial stellar rotation period $P_\mathrm{start} = 10$~days \citep[comparable to the \textit{slow} models in][]{Amard19}. 
During the initial stellar evolution ($t<t_\mathrm{0}$), the rotation period is fixed to 10~days. We note that the rotation period does change during the first 2~Myr. The respective effects, however, are beyond the scope of this work and will be treated in subsequent studies.

In order to successfully start a simulation combining stellar and disk evolution, initial models, which represent a T~Tauri star and its disk at an age of $t_\mathrm{0}$ have to be calculated.
The initial models are based on the aforementioned stellar and disk parameters.

\subsection{Initial stellar evolution: $t<t_{\rm 0}$}\label{sec:star_init}

\begin{figure}
    \centering
         \resizebox{\hsize}{!}{\includegraphics{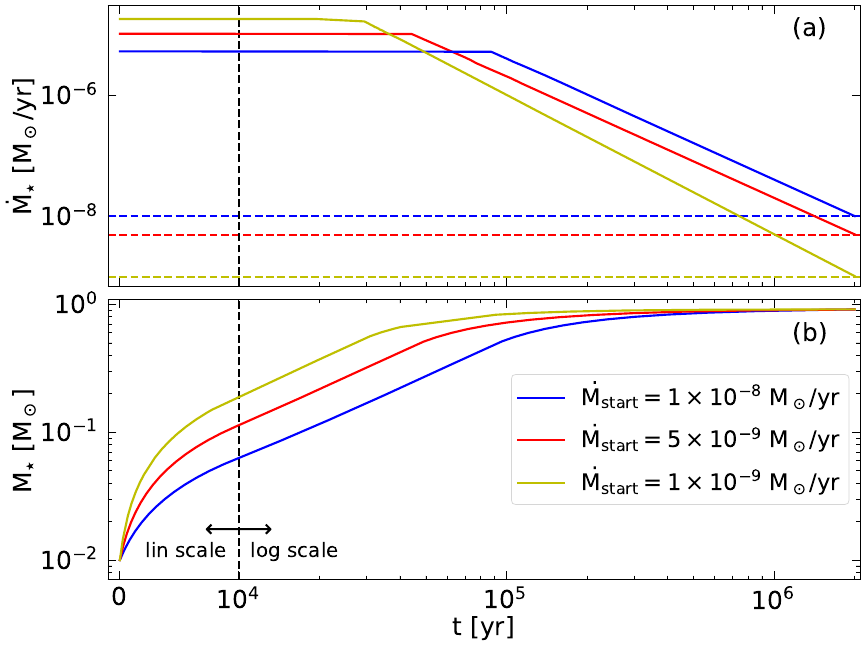}}
    \caption{Initial stellar evolution for $t<t_{\rm 0}$. Panel (a): Accretion history of the star. Initially, the star accretes with a constant accretion rate $\Dot{M}_{\rm init}$. At time $t_{\rm dec}$, the accretion rate drops according to $\Dot{M}_\star\propto t^{-2}$ (see text). The colored dashed lines show the respective value of $\Dot{M}_{\rm start}$.
    Panel (b): The evolution of the stellar mass. Starting from the initial mass of $0.01~\mathrm{M_\odot}$, all models reach their final mass of $0.91~\mathrm{M_\odot}$ after 2~Myr. 
    The yellow, red and blue lines show models with $\Dot{M}_{\rm start} = [1\times10^{-9},5\times10^{-9},1\times10^{-8}]~\Msunpyr$, respectively.
    To show the entire initial stellar evolution, starting from $t=0$~yr, we use the \textit{symlog} representation \citep[][]{Hunter2007}. The vertical dashed line indicates the transition from the linear scale (for $t\leq10^4$~yr) to the logarithmic scale (for $t>10^4$~yr).
    }
    \label{fig:star_h_accr}
\end{figure}

The early evolution of a protostar depends on the temporal evolution of the accretion rate \citep[e.g.][]{Kunitomo17, Steindl21}.
Unfortunately, the accretion history of a young star is difficult to measure, hardly constrained, and, thus, treated as a free parameter in our model. 
Based on the numerical results by \cite{Vorobyov17c}, we start with a constant accretion rate $\Dot{M}_{\rm init}$ followed by a decrease towards our desired value of $\Dot{M}_{\rm start}$ according to $\Dot{M}_\star\propto t^{-2}$ (see \fig{fig:star_h_accr}). 
We note that for the models with $\Dot{M}_{\rm start} =1\times10^{-9}~\Msunpyr$, we use $\Dot{M}_\star\propto t^{-2.3}$ to avoid values for $\Dot{M}_{\rm init}>2\times10^{-5}~\Msunpyr$.
The reason for this variation is convergence problems in the MESA initial model with higher initial accretion rates \citep[e.g., Sec.~5.1.5 in][]{Steindl21}.
The values of $\Dot{M}_{\rm init}$ and the time, at which the accretion rate begins to decrease, $t_{\rm dec}$, depend on $\Dot{M}_{\rm start}$ as well as the desired stellar mass at $t_{\rm 0}$.
For a starting accretion rate of $\Dot{M}_{\rm start} = [1\times10^{-9},5\times10^{-9},1\times10^{-8}]~\Msunpyr$, we find $\Dot{M}_{\rm init} = [1.87\times10^{-5},1.03\times10^{-5},5.29\times10^{-6}]~\Msunpyr$ and $t_{\rm dec} = [28,45,87]$~kyr, respectively.
We note that the accretion history of the star is identical for both metallicities.
We present our results for the initial stellar evolution in three parts: from the stellar seed to the transition towards the Hayashi track corresponding to an evolution time $\tau_\mathrm{H}$, from $\tau_\mathrm{H}$ along the Hayashi track towards $t_\mathrm{0}$ and finally reaching $t_\mathrm{0}$, which marks the starting point for our combined stellar and disk model.

\begin{figure}
    \centering
         \resizebox{\hsize}{!}{\includegraphics{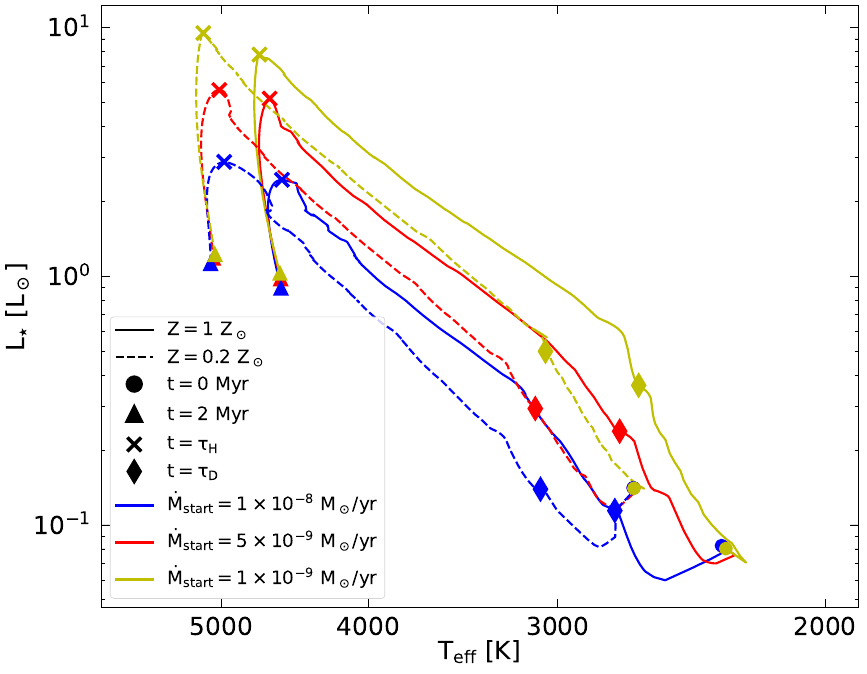}}
    \caption{
    Evolution on the HRD for $t<t_{\rm 0}$ for both metallicities $Z=1~\mathrm{Z_\odot}$ (solid lines) and $Z=0.2~\mathrm{Z_\odot}$ (dashed lines). The values at time $t=0$~Myr ($t=2$~Myr) are marked with circles (triangles) and the yellow, red and blue lines show models with $\Dot{M}_{\rm start} = [1\times10^{-9},5\times10^{-9},1\times10^{-8}]~\Msunpyr$, respectively.
    The onset of Deuterium burning ($t=\tau_\mathrm{D}$) is marked with a diamond.
    The transition from expansion towards the Hayashi track ($t = \tau_\mathrm{H}$) is marked with a cross.
    }
    \label{fig:star_h_hrd}
\end{figure}

\textit{From the stellar seed to $\tau_\mathrm{H}$:}
Considering the evolution of the star on a Hertzsprung-Russell-Diagram (HRD) for $t<t_{\rm 0}$, it is noticeable that stellar metallicity and the accretion history have a distinct influence on the star (see \fig{fig:star_h_hrd}).
Stars with lower metallicity are usually more compact and hotter compared to higher metallicities and their position on the HRD is moved to the upper left. 
The accretion history influences the amount of accretion energy added to the stellar interior (see \sref{sec:MESA} and \equo{eq:beta_mesa}).
In our model, the fraction of added energy depends on the magnitude of the accretion rate.
A higher accretion rate $\Dot{M}_{\rm init}$ results in a higher amount of energy added to the star.\footnote{The different behavior at the beginning of the evolution (the first $\sim 100$~years) can be explained by the outer boundary condition of the MESA code that reacts differently on different values of $L_\mathrm{add}$, which depends on the accretion rate.}
We note that a higher $\Dot{M}_{\rm init}$ results in a lower $\Dot{M}_{\rm start}$ (see \fig{fig:star_h_accr}).
The additional energy causes the star to expand and the stellar luminosity increases with an increasing amount of energy added to the star.
On the HRD, the stars are shifted in a vertical direction dependent on their accretion history (see \fig{fig:star_h_hrd}).
In other words, for a given effective temperature and luminosity, a star is younger if less energy is added to the star during its formation \citep[e.g.][]{Kunitomo17}.
At a few kyr, Deuterium burning sets in ($\tau_\mathrm{D}$, marked with diamonds in \fig{fig:star_h_hrd}) and additionally increases the stellar luminosity. 
The onset times range between 1.31~kyr and 2.72~kyr (see \tab{tab:hayashi}). 
The point of maximum luminosity ($\tau_\mathrm{H}$, marked with crosses in \fig{fig:star_h_hrd}) indicates a transition. 
Before $\tau_\mathrm{H}$, the pre-MS star derives most of its energy from accretion and Deuterium burning and afterward from contraction on the Kelvin-Helmholtz timescale \citep[e.g.,][]{Stahler2004}.
The young star turns towards the Hayashi track.

\textit{Evolution on the Hayashi track:}
As the accretion rate decreases over time (see \fig{fig:star_h_accr}), the amount of energy added to the stellar interior decreases as well and the expansion of the radius stops.
The star contracts subsequently on the Hayashi track, similar to the classical stellar evolution model.
In \fig{fig:star_h_hrd}, this transition towards the Hayashi track is marked with a cross ('x').
The respective stellar mass ($M_\mathrm{H}$) and evolution time ($\tau_\mathrm{H}$) is summarized in \tab{tab:hayashi}. 
The stellar masses ($M_\mathrm{H}$) range between 0.66 and 0.83~$\mathrm{M_\odot}$ (with slightly lower values for $Z=0.2~\mathrm{Z_\odot}$).
The time $\tau_\mathrm{H}$ depends, as expected, on the accretion history.
Shortly after the accretion rate decreases from $\Dot{M}_\mathrm{init}$ towards $\Dot{M}_\mathrm{start}$ the amount of energy added to the star is insufficient to expand the radius and counteract contraction.
Our models are comparable with the results of \cite{Kunitomo17} for warm/hot accretion models.

\begin{figure}
    \centering
         \resizebox{\hsize}{!}{\includegraphics{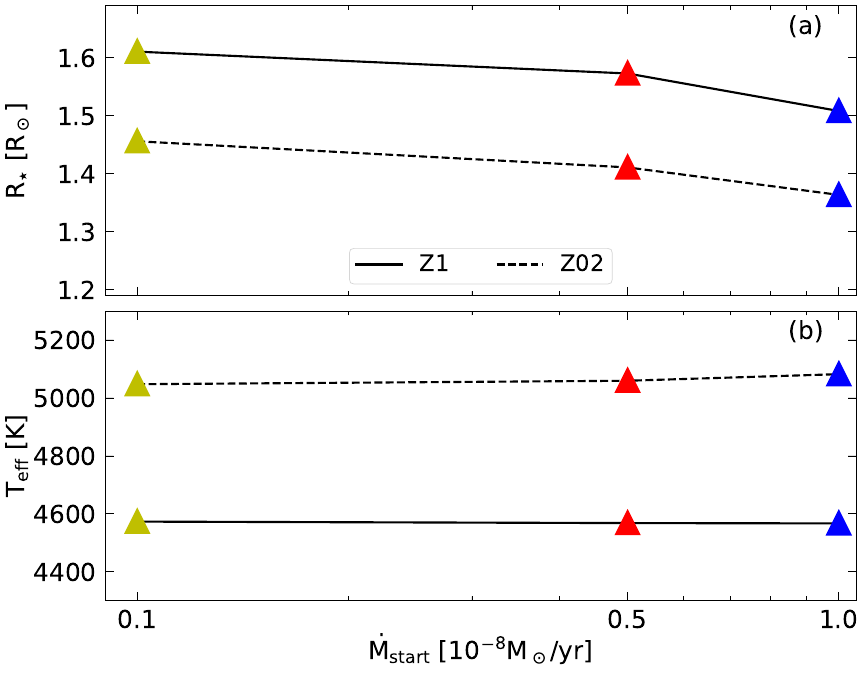}}
    \caption{
    Stellar radii and temperatures at $t_\mathrm{0}=2$~Myr for $\Dot{M}_\mathrm{start}= 5\times 10^{-9}$ (red triangles) and $1 \times 10^{-8}$ (blue triangles). Solar (sub-solar) metallicity values are connected by a solid (dashed) line, respectively.
    }
    \label{fig:star_init}
\end{figure}

\textit{Reaching $t_\mathrm{0}=2$~Myr:}
When reaching an age of $t_\mathrm{0}=2$~Myr, stellar radii and temperatures (see \fig{fig:star_init}) are used to calculate the initial disk model (see \sref{sec:disk_init}).
While the stellar radius differs due to the accretion history, the effective temperature stays approximately constant.
When connecting the colored triangles in \fig{fig:star_h_hrd} for $Z=1~\mathrm{Z_\odot}$, they are located on a vertical line indicating evolution on the Hayashi track (stellar luminosity decreasing with age).
For $Z=0.2~\mathrm{Z_\odot}$ however, the stars at $t_\mathrm{0}$ already turn towards the Henyey track (stellar luminosity and effective temperature increase with age) indicating the development of a radiative core.

\begin{table}[ht]
\centering
\caption{Onset of Deuterium burning ($\tau_\mathrm{D}$), stellar mass ($M_\mathrm{H}$) and evolution time ($\tau_\mathrm{H}$) the star turns onto the Hayashi track}        
\begin{tabular}{c c c c c}         
\hline\hline 
$\Dot{M}_\mathrm{start}$ [$\Msunpyr$] &  $Z~\mathrm{[Z_\odot]}$ & $M_\mathrm{H}~\mathrm{[M_\odot]}$ & $\tau_\mathrm{H}~\mathrm{[Myr]}$ & $\tau_\mathrm{D}$~[kyr] \\
\hline
$1\times 10^{-9}$  & 1   & 0.82 & 0.09 & 1.45 \\
$5\times 10^{-9}$  & 1   & 0.71 & 0.10 & 2.31 \\
$1\times 10^{-8}$  & 1   & 0.69 & 0.16 & 2.72\\
$1\times 10^{-9}$  & 0.2 & 0.81 & 0.09 & 1.31\\
$5\times 10^{-9}$  & 0.2 & 0.66 & 0.08 & 1.84\\
$1\times 10^{-8}$  & 0.2 & 0.66 & 0.15 & 2.59\\

\hline\hline                                            
\end{tabular}
\label{tab:hayashi}  
\end{table}

\subsection{Initial disk model}\label{sec:disk_init}

\begin{figure}
    \centering
         \resizebox{\hsize}{!}{\includegraphics{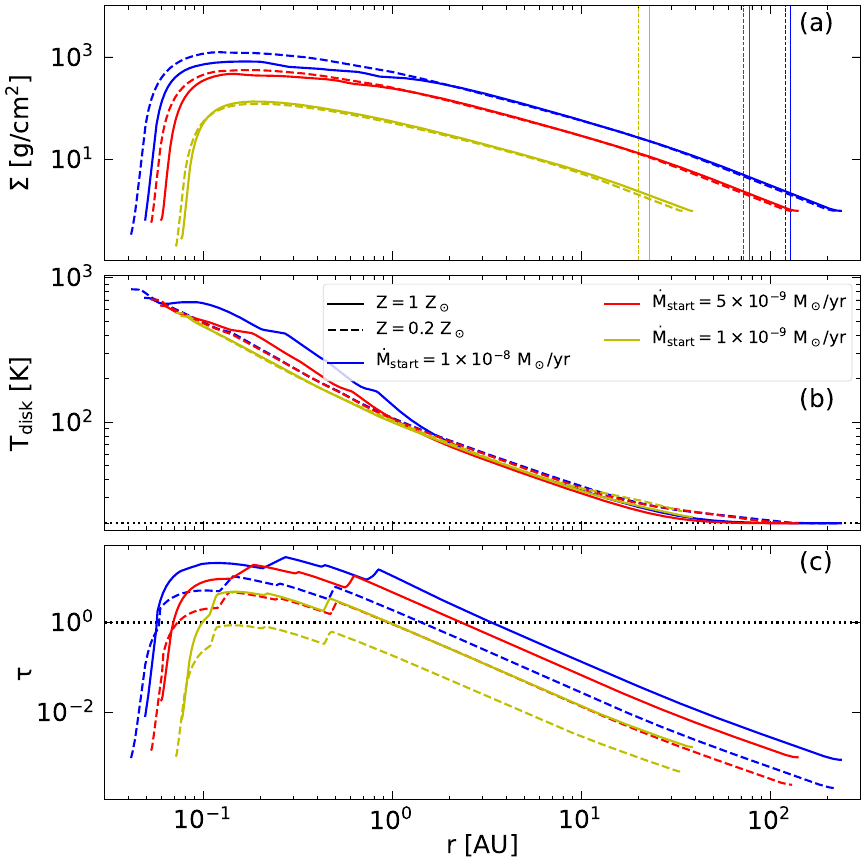}}
    \caption{
    The radial profile of the initial disk model for both metallicities $Z=1~\mathrm{Z_\odot}$ (solid lines) and $Z=0.2~\mathrm{Z_\odot}$ (dashed lines). The yellow, red and blue lines indicate models with $\Dot{M}_{\rm start} = [1\times10^{-9},5\times10^{-9},1\times10^{-8}]~\Msunpyr$, respectively.
    Panel (a) shows the disk surface density $\Sigma$, Panel (b) the disk midplane temperature $T_\mathrm{disk}$, and Panel (c) the optical depth $\tau$, respectively. 
    The thin vertical lines in Panel (a) indicate the respective position of $R_\mathrm{c}$.
    The horizontal line in Panel (b) indicates the ambient temperature $T_\mathrm{amb} = 20$~K.
    The horizontal dashed line in Panel (c) symbolizes $\tau = 1$, dividing optical thin regions ($\tau << 1$) from optical thick regions ($\tau >> 1$).
    }
    \label{fig:disk_init}
\end{figure}

Starting a simulation with an implicit method requires an initial model that already solves \equs{eq:cont}{eq:ene} \citep[e.g.][]{Dorfi2006}. 
Following \cite{Steiner21}, we construct a steady-state accretion disk that serves as the initial model in our simulation. 
The mass flow rate throughout the disk corresponds to the respective value of $\Dot{M}_\mathrm{start}$.
In \fig{fig:disk_init}, the radial surface density (Panel~a), disk midplane temperature (Panel~b), and the optical depth $\tau$ (Panel~c) profile of the initial disk models are shown.
The stellar metallicities are represented by solid lines ($Z=1~\mathrm{Z_\odot}$) and dashed lines ($Z=0.2~\mathrm{Z_\odot}$) respectively.
The yellow, red and blue lines symbolize models with $\Dot{M}_{\rm start} = [1\times10^{-9},5\times10^{-9},1\times10^{-8}]~\Msunpyr$, respectively.

For larger initial accretion rates, the disks are more massive, have a higher temperature, and are pushed closer towards the star \citep[e.g.][]{Romanova02, bessolaz08, Romanova14, Steiner21}.
Due to the smaller stellar radii at $t_\mathrm{0}$ (see \fig{fig:star_init}), the absolute values of the inner disk edge $r_\mathrm{trunc}$ is smaller for $Z=0.2~\mathrm{Z_\odot}$.
The radial temperature profile can be divided into three regions. 
Close to the star the disk is directly heated by stellar irradiation $r\lesssim 10^{-1}$.
For low metallicities, stellar luminosity is larger and, thus, the disk temperature is higher.
From $r\lesssim 10^{-1}$~AU to $\sim 1$~AU, the midplane of the disk is shielded from stellar irradiation and the disk is optically thick (Panel~c in \fig{fig:disk_init}).
This region is referred to as the optical thick zone, $\mathrm{R_{thick}}$, from now on.
In this region, the disk midplane temperature ranges from $\sim 100$ to 1000~K and is dominated by dust opacities \citep[e.g.,][]{Semenov2003}. 
In our model, we assume that the disk gas-to-dust ratio corresponds to the stellar metallicity, $f_\mathrm{dust} = 0.014$ and $0.0028$ for $Z = 1~\mathrm{Z_\odot}$ and $0.2~\mathrm{Z_\odot}$, respectively.
Consequently, larger stellar metallicity results in higher optical depths and temperatures in the disk midplane compared to lower stellar metallicities.
We note that in the models with $\Dot{M}_\mathrm{start}=1\times 10^{-9}\Msunpyr$, these effects are almost negligible due to the low surface density values.
Outside $R_\mathrm{thick}$, the disk temperature is again dominated by stellar irradiation and smaller stellar metallicities result in higher disk temperatures.
The disk temperature decreases with increasing disk radius until the ambient temperature of $T_\mathrm{amb}= 20$~K is reached (see Panel b in \fig{fig:disk_init}).

To compare the initial disk models to observations and other theoretical models, we summarized the initial disk sizes, given by $R_\mathrm{c}$, and masses in \tab{tab:init_disk}. 
The characteristic radius of the disk $R_\mathrm{c}$ includes $2/3$ of the disk's mass.
The values of $R_\mathrm{c}$, ranging between 20~AU and 128~AU, and $m_\mathrm{disk}$, ranging from 0.001~$\mathrm{M_\odot}$ to 0.055~$\mathrm{M_\odot}$ are within the limits of observed disk radii and masses in Ophiuchus~II \citep[][]{Andrews2010} and Lupus star-forming regions \citep[][]{Ansdell2018}.

\begin{table}[ht]
\centering
\caption{
Characteristic radius $R_\mathrm{c}$ and disk mass $m_\mathrm{disk}$ of the initial disk model
}     
\begin{tabular}{c c c c}         
\hline\hline 
$\Dot{M}_\mathrm{start}$ [$\Msunpyr$] &  $Z~\mathrm{[Z_\odot]}$ & $R_\mathrm{c}$~[AU] & $m_\mathrm{disk}$~[$\mathrm{M_\odot}$] \\
\hline
$1\times 10^{-9}$  & 1   & 23 & 0.003 \\
$1\times 10^{-9}$  & 0.2   & 20 & 0.001 \\
$5\times 10^{-9}$  & 1   & 78 & 0.019 \\
$5\times 10^{-9}$  & 0.2   & 72 & 0.017 \\
$1\times 10^{-8}$  & 1   & 128 & 0.055 \\
$1\times 10^{-8}$  & 0.2   & 121 & 0.051 \\

\hline\hline                                            
\end{tabular}
\label{tab:init_disk}  
\end{table}

\subsection{Disk evolution: $t>t_{\rm 0}$}\label{sec:disk_evo}

\begin{figure}
    \centering
         \resizebox{\hsize}{!}{\includegraphics{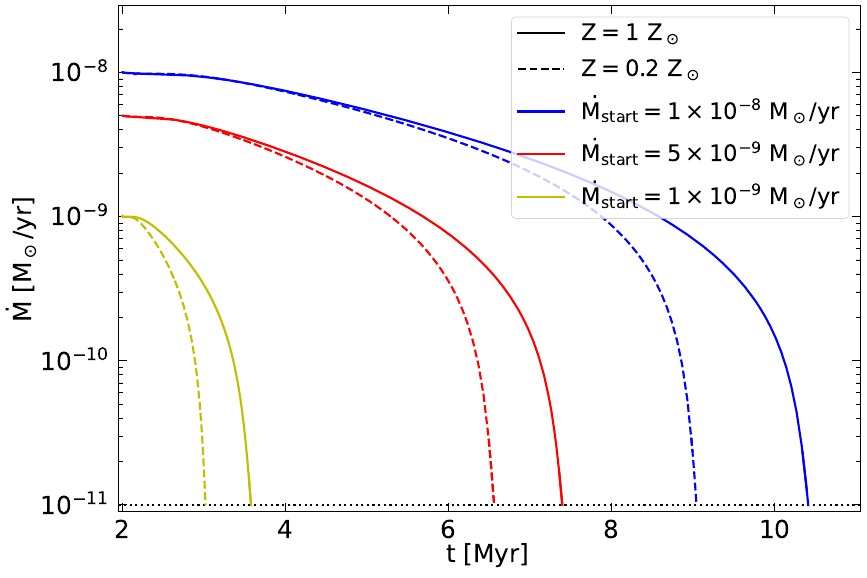}}
    \caption{
    Accretion rate onto the star over time for both metallicities $Z=1~\mathrm{Z_\odot}$ (solid lines) and $Z=0.2~\mathrm{Z_\odot}$ (dashed lines). The yellow, red and blue lines indicate models with $\Dot{M}_{\rm start} = [1\times10^{-9},5\times10^{-9},1\times10^{-8}]~\Msunpyr$, respectively. The black dotted line shows the termination criterion used in our simulations of $\Dot{M} = 10^{-11}~\Msunpyr$.
    }
    \label{fig:disk_evo}
\end{figure}

Starting from the stellar and disk initial models, we evolve the star-disk system until the accretion rate onto the star decreases below $10^{-11}\Msunpyr$.
In \fig{fig:disk_evo}, the evolution of the disk accretion rate is shown for both metallicities $Z=1~\mathrm{Z_\odot}$ (solid lines) and $Z=0.2~\mathrm{Z_\odot}$ (dashed lines). 
The red and blue lines indicate models with $\Dot{M}_{\rm start} = [1\times10^{-9},5\times10^{-9},1\times10^{-8}]~\Msunpyr$, respectively.
Independent of the accretion history of the star, a lower stellar metallicity results in a shorter disk lifetime\footnote{In this context, disk lifetime is the time, at which the accretion rate has dropped to $10^{-11}\Msunpyr$.}.

Observations of disk fractions in low metallicity clusters \citep[][]{Yasui10} and statistical studies \citep[][]{Elsender2021} suggest that the disk lifetime is indeed shorter for lower stellar metallicities. 
\cite{Nakatani18} explain this behavior with a metallicity-dependent photo-evaporation rate.
As we did not include photo-evaporation in our model, we are dealing with a different mechanism affecting the disk lifetime.

Owing to the higher stellar luminosity at lower metallicities, the innermost disk region (inside $R_\mathrm{thick}$) and the outer disk regions (outside $R_\mathrm{thick}$) are hotter compared to higher metallicities (see \fig{fig:disk_init}).
With higher disk temperatures, the disk's viscosity $\nu$ increases.
Thus, the viscous timescale $\tau_\mathrm{\nu} = r^2 / \nu$ decreases.
Since the viscous timescale is a measure of the time, in which the disk material is accreted onto the star, the disk material accretes faster onto the star for lower metallicities.
As shown in \fig{fig:star_evo_late}, in which the evolution of the stellar parameters for $t\geq t_\mathrm{0}$ are summarized, the differences in \textbf{intrinsic} luminosity between solar and sub-solar metallicity increase for older ages.
As a consequence, the older a star-disk system is, the greater the difference in stellar irradiation and disk heating between solar and sub-solar metallicities. 
Usually, the (global) viscous timescale is measured in the outer ($\sim 10$~AU) disk regions \citep[e.g.,][]{armitage11}.
During the past years, however, recent studies highlighted the importance of the inner disk, the sub-AU region \citep[e.e.g,][]{Vorobyov19, Hennebelle2020}. 
The accretion rate onto the star is eventually regulated by the interaction between the star and the disk.

\begin{figure}
    \centering
         \resizebox{\hsize}{!}{\includegraphics{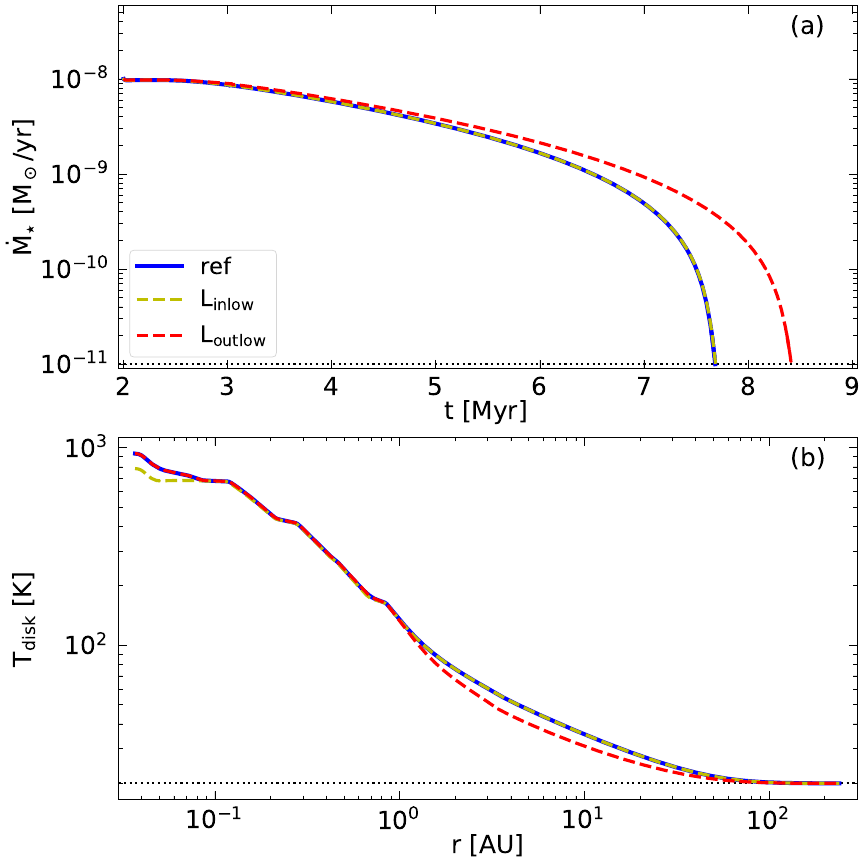}}
    \caption{
    Comparison between models $\mathrm{L_{inlow}}$ (yellow line) and $\mathrm{L_{outlow}}$ (red line). 
    The reference model (ref, blue line) corresponds to a $Z = 1~\mathrm{Z_\odot}$ star with $\Dot{M}_\mathrm{start}= 1\times 10^{-8}~\Msunpyr$.
    The stellar luminosity is artificially reduced by 25~\% in the inner 1~AU ($\mathrm{L_{inlow}}$) and outside 1~AU ($\mathrm{L_{outlow}}$).
    Panel (a) shows the evolution of the accretion rate. The black dotted line shows the termination criterion used in our simulations of $\Dot{M} = 10^{-11}~\Msunpyr$.
    Panel (b) shows the temperature structure of the initial disk model. 
    For model $\mathrm{L_{inlow}}$, the inner disk region is cooler and for model $\mathrm{L_{outlow}}$, the outer disk region is cooler compared to the reference model. The black dotted line shows the ambient temperature of 20~K.
    }
    \label{fig:Ltot_test}
\end{figure}

To test, which region (or perhaps even both) influences the disk lifetime, we artificially shield a certain part of the disk from 25~\% of the stellar irradiation, resulting in two test models with a lower disk temperature in the inner disk region $<1$~AU (model $\mathrm{L_{inlow}}$) and the outer disk region $> 1$~AU ($\mathrm{L_{outlow}}$), respectively (see Panel b of \fig{fig:Ltot_test}).
We compare the disk evolution of the models $\mathrm{L_{inlow}}$ and $\mathrm{L_{outlow}}$ to a reference case (ref) without shielding certain regions from stellar irradiation.
The evolution of the accretion rate (Panel a in \fig{fig:Ltot_test}) clearly shows that the different disk temperature in the outer region influences the disk lifetime.
We note that in our current model, short-term effects such as episodic outbursts are not included.
The number and strength of these events depend strongly on the inner disk region \citep[e.g.,][]{Vorobyov20, Steiner21} and can also affect the long-term disk evolution.
Including a more detailed physical model, could increase the importance of the inner disk region.
In the simulation shown in \fig{fig:Ltot_test}, the stellar evolution has been switched off to preserve computational resources.
The qualitative statement of the comparison should not be impaired by this choice.

\subsection{Influence of metallicity on stellar rotation}\label{sec:rot_evo}

The influence of metallicity on stellar rotation has been studied for older ($\gtrsim 1$~Gyr) stellar populations based on the dependence of wind-breaking formulations on stellar metallicity \citep[][]{Amard20,Amard20b}.
Low metallicity stars spin faster compared to their solar metallicity counterparts.
During the disk phase, however, mechanisms besides wind-braking influence the stellar rotation rate (see \sref{sec:stellar_spin_model}), and these studies assume a constant stellar period (disk-locking).

\begin{figure}
    \centering
         \resizebox{\hsize}{!}{\includegraphics{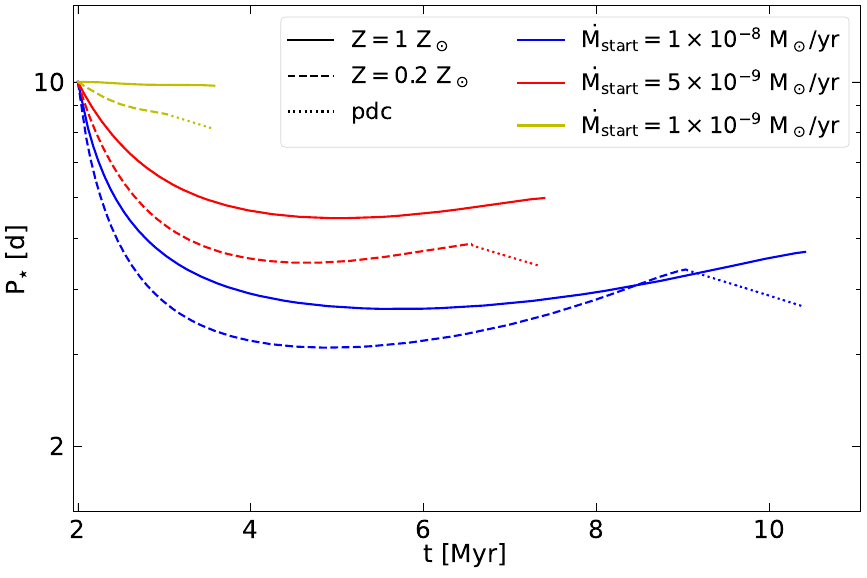}}
    \caption{
    Stellar rotational periods over time for both metallicities $Z=1~\mathrm{Z_\odot}$ (solid lines) and $Z=0.2~\mathrm{Z_\odot}$ (dashed lines). The yellow, red and blue lines indicate models with $\Dot{M}_{\rm start} = [1\times10^{-9},5\times10^{-9},1\times10^{-8}]~\Msunpyr$, respectively. For low metallicities, stellar spin-up due to contraction (post disk contraction, pdc) until the disk lifetime of the solar metallicity is reached, is indicated by colored dotted lines. 
    }
    \label{fig:spin_evo}
\end{figure}

In \fig{fig:spin_evo}, the evolution of the stellar rotation period is shown over time for both metallicities $Z=1~\mathrm{Z_\odot}$ (solid lines) and $Z=0.2~\mathrm{Z_\odot}$ (dashed lines). 
The yellow, red and blue lines indicate models with $\Dot{M}_{\rm start} = [1\times10^{-9},5\times10^{-9},1\times10^{-8}]~\Msunpyr$, respectively.
All simulations are started at a stellar period at $t_\mathrm{0}$ of $P_\mathrm{0} = 10$~days. 
We expect that a variation in $P_\mathrm{0}$ does not change the qualitative result shown in \fig{fig:spin_evo}, but rather shifts the respective lines in the vertical direction.
Independent of the stellar accretion history, stars with lower stellar metallicities rotate faster compared to solar metallicity values. 
We can identify three reasons for this development during the disk phase.

\textit{Stellar radius:} First of all, the metallicity-dependent stellar initial values have an effect on spin evolution. Low metallicities result in smaller stellar radii at $t_\mathrm{0}$. 
According to the stellar spin model described in \sref{sec:stellar_spin_model}, the two mechanisms that can remove angular momentum from the star (APSW and MEs) both scale with the stellar radius.
Consequently, smaller stellar radius results in less effective stellar spin-down mechanisms and low metallicity stars spin faster.

\textit{Stellar contraction:} Contraction of the pre-MS star causes the star to spin up. Dependent on the stellar metallicity and the accretion history of the star, the evolutionary stage at a certain time and, thus, the stellar contraction can differ (see \fig{fig:star_h_hrd}).
Panel (a) in \fig{fig:star_evo_late} shows the evolution of the stellar radii over time for both metallicities $Z=1~\mathrm{Z_\odot}$ (solid lines) and $Z=0.2~\mathrm{Z_\odot}$ (dashed lines).
For low metallicities, the star contracts more slowly. 
A slower stellar contraction rate counteracts the effect of smaller stellar radii with respect to stellar rotation (see above).

\textit{Different disk lifetime:} For low metallicities, the lifetime of an accretion disk is, within the scope of our models, $\sim 1$~Myr shorter (see \sref{sec:disk_evo}).
After the disk is dissolved, there are no mechanisms that can remove angular momentum and the star spins up due to contraction.
\footnote{We note that there is also a stellar wind mechanism acting after the disk phase. Compared to stellar contraction, however, its effect is small and can be neglected during a relatively short time period of $\sim 1$~Myr. After contraction has ended and the star evolves towards and on the MS, stellar winds slow down the star on longer timescales \citep[e.g.,][]{Gallet13}. }
Stellar spin-up due to contraction, without the interference of a disk, starts at an earlier age. 
As a result, the difference in rotation period between low and solar metallicity stars increases further.

\subsection{Robustness of the results}

The results presented in this section are based on certain, fixed stellar and disk parameters, for example, the stellar magnetic field strength $B_\star$ and the disk's viscous $\alpha$-parameter.
We want to verify the robustness of our results by changing these parameters.
For our models with $\Dot{M}_{\rm start} = [1\times10^{-9},5\times10^{-9},1\times10^{-8}]~\Msunpyr$, the stellar magnetic field is reduced to 0.5~kG (model MC1), the initial rotation period is set to 2~days (model MC2), and the viscous $\alpha$-parameter is reduced to 0.005 (model MC3), respectively.

\begin{figure*}
    \centering
         \resizebox{\hsize}{!}{\includegraphics{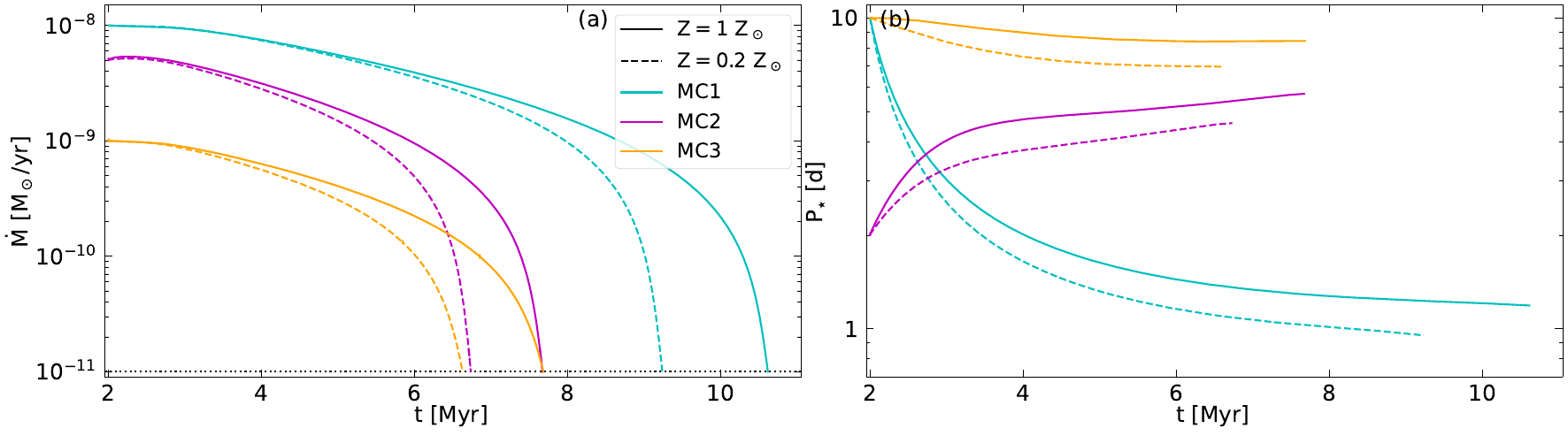}}
    \caption{
    Influence of the stellar magnetic field strength $B_\star$, initial rotation period, and viscous $\alpha$-parameter on the results for both metallicities $Z=1~\mathrm{Z_\odot}$ (solid lines) and $Z=0.2~\mathrm{Z_\odot}$ (dashed lines). The cyan, magenta, and orange lines show models MC1, MC2, and MC3, respectively.
    Panel (a): Same as \fig{fig:disk_evo}. Panel (b): Same as \fig{fig:spin_evo} without the post-disk contraction (pdc).
    }
    \label{fig:verify}
\end{figure*}

\fig{fig:verify} shows the evolution of the accretion disk (Panel~a, see \fig{fig:disk_evo}) and the stellar spin evolution (Panel~b, see \fig{fig:spin_evo}) for models MC1, MC2, and MC3.
Similar to the results shown in \sref{sec:disk_evo} and \sref{sec:rot_evo}, the trends regarding disk lifetime and stellar rotation period with respect to the metallicity are clearly visible.
This indicates that our results are valid over a wider range of parameters.

%
%
%
%
%
%



\section{Discussion}
\label{sec:discussion}

\subsection{Comparing our results to observed rotation periods}\label{sec:comp_rot}

At ages, $\gtrsim 10$~Myr, the rotation periods of young stars can directly affect the high energy radiation that interacts with planetary atmospheres \citep[e.g.,][]{France18}.
Fast rotating stars usually have stronger high energy emissions compared to slow rotating stars \citep[e.g.,][]{Johnstone2021}. 
Our results indicate that low-metallicity stars rotate faster than their solar-metallicity counterparts after the disk phase (with differences up to 2~days, see \sref{sec:rot_evo}).

We compare our results to observed periods in young clusters to check the plausibility of the faster rotation periods for low-metallicity stars.
Unfortunately, the age estimates of young clusters are subjected to large uncertainties, depending on the stellar evolutionary tracks used to fit the data.
Additionally, within one cluster, the different stellar populations can co-exist expanding the age span within one cluster up to several Myr.
With these limitations in mind, we choose 6 young clusters with ages between $\sim 2$~Myr and $\lesssim 5$~Myr.
The clusters used in this comparison are NGC~6530, NGC~2362, NGC~2264, Orion star-forming cluster (OSFC), $\sigma$~Ori and Taurus star-forming region with observed rotation periods taken from \cite{Henderson12}, \cite{Irwin08}, \cite{Affer13}, \cite{Serna2021}, \cite{Cody2010} and \cite{Rebull2020}, respectively. 
The observed ages, mass ranges, and metallicities of these clusters are summarized in \tab{tab:comp_rot}.
We note that our models are not included in the mass range of observed periods in $\sigma$~Ori and low-mass stars tend to rotate faster compared to higher stellar masses \citep[e.g.,][]{Irwin08}. Owing to the small number of available data, we choose to include $\sigma$~Ori in this comparison.

\begin{figure}
    \centering
         \resizebox{\hsize}{!}{\includegraphics{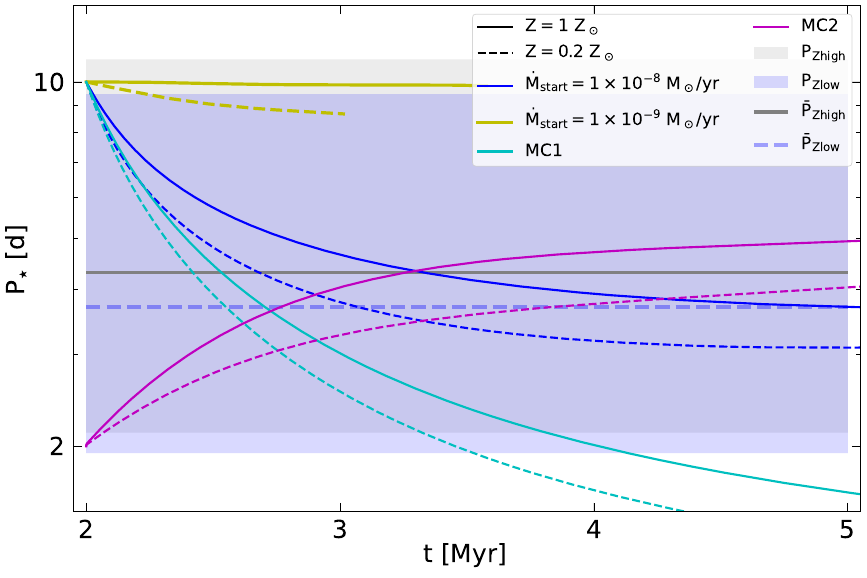}}
    \caption{
    Stellar rotational periods over time for both metallicities $Z=1~\mathrm{Z_\odot}$ (solid lines) and $Z=0.2~\mathrm{Z_\odot}$ (dashed lines). The colored lines represent the respective model. 
    The ranges of the observed periods of young clusters with ages between $\sim 3$~Myr and $\lesssim 5$~Myr (see text) are indicated by the light grey (light blue) area for $\mathrm{P_{Zhigh}}$ ($\mathrm{P_{Zlow}}$), respectively. The mean values ($\mathrm{\Bar{P}}$) are indicated by horizontal lines in the respective color.
    }
    \label{fig:spin_comp}
\end{figure}

The observed rotation periods are combined in two bins: the cluster metallicity is assumed to be solar or above ($\mathrm{P_{Zhigh}}$) and the metallicity is assumed to be below the solar value ($\mathrm{P_{Zlow}}$). 
In \fig{fig:spin_comp}, the ranges of observed rotation periods from the $25^{\mathrm{th}}$ to the $90^{\mathrm{th}}$ percentile are shown for $\mathrm{P_{Zhigh}}$ (light grey) and $\mathrm{P_{Zlow}}$ (light blue). The mean values ($\mathrm{\Bar{P}}$) are indicated by horizontal lines in the respective color.
Consistent with our results, the range of observed periods is shifted to faster values for lower metallicities.
The difference in rotation period due to lower metallicity values, however, is small, compared to other stellar parameters, e.g., the stellar magnetic field strength. 
We note that the lower stellar mass range of observed periods in $\sigma$~Ori could further reduce the difference even further.
Thus, the faster rotation periods due to lower metallicity values alone, will probably not affect the stellar high energy output and, in consequence, the planetary atmosphere significantly.
If, on the other hand, the magnetic field strength (that is assumed to be constant during our simulations) can be related to metallicity, the importance of metallicity could increase. 
For such relations, however, additional theoretical and observational work is needed.

\subsection{Early evolution of the star-disk system: Episodic accretion during $t<t_\mathrm{0}$}\label{sec:early_evo}
In this study, we start the combined star-disk simulation at $t_\mathrm{0}=2$~Myr.
For ages $t<t_\mathrm{0}$ fixed parameters are used to define the accretion history on the star (see \fig{fig:star_h_accr}).
The accretion rate for $t<t_\mathrm{0}$, however, can vary significantly.
Episodic accretion events (outbursts) triggered by gravitational instabilities and disk fragmentation, thermal and magneto-rotational instabilities, or a combination of those, can increase the accretion rate by orders of magnitude over short periods of time \citep[e.g.][]{nayakshin12, Bae14,Vorobyov15, Vorobyov17c}.

The amount of energy added to the stellar interior depends on the accretion rate. 
Stronger and more numerous outbursts result in more energy added to the star.
The number and intensity of outbursts also depend on the stellar mass \citep[e.g.,][]{Vorobyov17c, Kadam2020}.
High stellar masses show more numerous and luminous outbursts compared to low-mass protostars.

Furthermore, the stellar rotation period is also affected by early evolution.
The principle contributions summarized in \sref{sec:stellar_spin_model} can influence the star starting from its formation.
It is interesting to note that during accretion outbursts the additional energy can cause the star to inflate \citep[e.g.,][]{Vorobyov17c}.
While the additional angular momentum carried by the accreted disk material spins up the star, the inflation of the stellar radius causes a spin-down effect.
We plan, for future studies, a starting point $t_\mathrm{0}$ reasonably close to the formation of the stellar seed, allowing the consideration of accretion outbursts and early stellar spin evolution.

\subsection{Angular momentum transport within the star}\label{sec:AM_within}

In this study, we assume solid body rotation for the star. 
This simplification is justified as long as the star is fully convective without the presence of a radiative core \citep[e.g.][]{Amard19}.
When a radiative core develops together with contraction on the pre-MS, a strong meridional circulation (flowing from North to South) transports angular momentum from the core towards the surface \citep[e.g.][]{Amard16,Amard19}.

In general, during its presence, a radiative core and the convective envelope above, which is still assumed to rotate as a solid body, have to be treated individually. 
The convective envelope is still affected by external torques plus the angular momentum flux induced by the meridional circulation. 
Additionally, the core and envelope can exchange angular momentum due to shear-induced turbulence \citep[e.g.][]{Amard19}.
The extent of this angular momentum transfer between the core and the surface of the star depends on the stellar rotation period and stellar mass.
Based on the calculations of \cite{Amard19}, stars rotate either close to solid-body rotation (valid for fast rotating stars during the first $\lesssim 10^{8}$~years) or the core rotates faster compared to the convective envelope. 

Observational confirmation of such a model poses problems. Asteroseismic data reveal only a few estimates of the core angular velocity of low-mass, main sequence stars \citep[][]{Benomar15}.
In contrast to the model results presented in \cite{Amard19}, these observations suggest slowly rotating stellar cores, motivating efforts to advance theoretical models and obtain more observational data.

On the other hand, helioseismic data suggest a fast-rotating solar core from g-mode observations \citep[][]{Fossat2017}.
In addition, precise TESS observations of the massive star HD~192575 reveal differential rotation between the near-core region and envelope \citep[][]{Burssens2021}. 
Based on their results, the near-core region rotates at a higher frequency.
To what extent these results can be applied to our model, however, is unclear due to the different structures and ages of the observed stars.
Furthermore, recent theoretical works of \cite{Buldgen2022} and \cite{Eggenberger2022} present progress in combining the theory of internal stellar angular momentum transport and observational constraint, for example, the solar surface lithium abundance.

\subsection{Evolution of $B_\star$}

The stellar magnetic field strength is a highly variable parameter that can change its magnitude within a short period of time \citep[several years, e.g.][]{Johnstone14}. 
The values used in this study for $B_\star$ must be understood as averaged over time.
Furthermore, the field topology is assumed to depend on the evolutionary stage \citep[e.g.][]{Zaire17,Emeriau17}.
With the presence of a radiative core, the magnetic field topology can become more complex, which can result in a weaker large-scale dipole component.
In our model, the magnetic field strength $B_\star$ represents the large-scale, dipole component of the magnetic field, which usually dominates the star-disk interaction region and the stellar spin evolution \citep[][]{Finnley18}.

For low metallicities, the star evolves further and even begins its turn towards the Henyey track, indicating the development of a radiative core.
During this time, the large-scale, dipole field strength can decrease, which affects the stellar spin evolution.
Referring to \fig{fig:verify}, such a decrease of $B_\star$ for low metallicity stars could be visualized by a gradual transition from the blue-dashed line (where $B_\star = 2.0$~kG is assumed) towards the cyan-dashed line ($B_\star = 0.5$~kG in model MC1).
As a result, the spread in rotational periods for different metallicity values increases further.

\subsection{Disk lifetimes in low metallicity clusters}

Our models predict a dependence of disk lifetimes on stellar metallicity. 
The disks around low metallicity stars dissolve faster compared to their solar metallicity counterparts (see \fig{fig:disk_evo}).
This correlation can be explained (within the scope of our model) by a hotter outer disk region enhancing viscosity, reducing the viscous timescale, and, thus, resulting in a shorter disk lifetime.

Over the last decade several observational studies can confirm this trend between metallicity and disk lifetime \citep[e.g.][]{Yasui10, Yasui16,Guarcello21, Yasui21}.
In general, the observations show a disk fraction in low metallicity disks that is lower compared to solar metallicity.
For example, at an age of 1~Myr, the disk fraction in a $Z=0.2~\mathrm{Z_\odot}$ cluster is $\sim 10$~\% \citep[e.g.][]{Yasui10} and considerably lower compared to solar metallicity values with disk fractions at 1~Myr of $60-80$~\%.

Different mechanisms affect the dispersal of the accretion disk in low-metallicity environments.
The ionization fraction in low metallicity disks can be higher compared to solar metallicities.
Smaller amounts of dust result in a slower recombination rate of free charges on dust grain surfaces.
A higher ionization fraction in the disk increases the efficiency of magneto-rotational instabilities \citep[][]{Balbus1991}, which drive accretion \citep[e.g.,][]{Hartmann09}.
Alternatively, the efficiency of photo-evaporation also depends on metallicity.
Lower metallicities and the resulting lower disk opacities allow deeper penetration of high energy radiation in the disk \citep[e.g.,][]{Ercolano10}.
As a result, a larger part of the disk is heated and the removal of gas and dust by photo-evaporation is facilitated.
More recently, numerical simulations \citep[e.g.,][]{Nakatani18} confirm the metallicity dependence of photo-evaporation. 
For sub-solar metallicities, the photo-evaporative mass-loss rates are higher and the accretion disk disperses at a faster rate.
It is worth noting that the observed disk lifetimes in low metallicity clusters cannot be conclusively explained by one mechanism alone \citep[e.g.,][]{Nakatani18} and results, more likely, from the combination of the aforementioned aspects.
In a more recent study, \cite{Elsender2021} proposes that increased stellar multiplicity in low-metallicity environments and resulting increased dynamical interactions might cause accretion disks to be smaller and have shorter lifetimes compared to higher metallicities.

\subsection{Differences between our results and observations}

The observations shown in \cite{Yasui2010b} suggest disk lifetimes of $\sim 3$~Myr for low-metallicity clusters, matching our results for $\Dot{M}_{\rm start} = 1\times10^{-9}~\Msunpyr$. 
In this context, cluster ages can be used to estimate disk lifetimes. 
The disk lifetimes for models with $\Dot{M}_{\rm start} = 5\times10^{-9}~\Msunpyr$ and $1\times10^{-8}~\Msunpyr$, on the other hand, are longer compared to observations by a factor of $\sim 3$. 
We want to address three possible reasons for this discrepancy.
\textit{First}, our models do currently not take photoevaporation into account. Including photoevaporation in future models will reduce our model's lifetime towards the observed values.
\textit{Second}, the cluster ages presented by \cite{Yasui2010b} are based on models that do not take magnetic fields into account. 
Cluster ages based on magnetic pre-MS stellar evolution models \citep[e.g.,][]{Feiden2015} are up to 2 Myr older compared to previous models \citep[e.g.,][]{Richert2018}. Disk lifetimes for low-metallicity stars could increase up to $\sim 5$~Myr.
\textit{Third}, the initial conditions in our models are independent of metallicity. 
It is, however, likely that the accretion rate, stellar rotation period, and stellar magnetic field strength depend on metallicity. Moving $t_\mathrm{0}$ to earlier evolutionary stages in combination with additional observational constraints for low-metallicity clusters will improve the quality of our results in future work.

\subsection{Previous spin evolution studies including metallicity}

Many new formulations have emerged over the last decade to explain the stellar spin evolution\citep[e.g.,][]{Matt15, Finnley18, Gallet19, Ireland21}.
These studies, however, are often based on solar properties and composition.
Thus, an explicit dependence on stellar metallicity is not given.

\cite{Amard19} presents, for the first time, a comprehensive set of stellar evolutionary tracks including rotation and different metallicities, using the \textit{STAREVOL} code.
They find that metallicity does influence the stellar spin evolution during the pre-MS and MS.
Low metallicity stars are more compact and evolve faster compared to their solar metallicity counterparts. Thus, during the pre-MS, they spin up faster and retain faster rotation periods during the MS.
Internally, low metallicity stars experience less differential rotation compared to solar metallicities.
A more recent study \cite{Amard20} provides an in-depth analysis of the effects of metallicity on different wind-braking formulations.
They also conclude that low-metallicity stars rotate faster than their solar metallicity counterparts. 
During the disk phase, a constant stellar period (disk-locking) is assumed for a fixed timescale independent of stellar mass and metallicity \citep[][]{Amard19, Amard20}.
This assumption is in conflict with observations of metallicity-dependent disk lifetimes \citep[e.g.,][]{Yasui10, Yasui16,Yasui21,Guarcello21}, photo-evaporation models \citep[e.g.,][]{Nakatani18} and the results presented in \sref{sec:disk_evo}.

Starting from the T-Tauri phase, our star-disk evolution model does include the effects of, e.g., metallicity, stellar magnetic field strength, stellar spin, and mass.
Over a wide range of stellar and disk parameters, we can provide realistic, metallicity-dependent disk lifetimes and stellar periods after the disk phase.
Those values can then be used as input parameters for the aforementioned spin evolution studies.

\subsection{Impact on disk chemistry}

Various studies confirm the influence of metallicity on the chemical evolution of disks. For example, by observing the chemical abundances of HII regions in 12 nearby dwarf galaxies, \cite{berg_carbon_2016} suggests that there are different chemistry cycles in non-solar metallicity environments. This evidence hints at the fact that disk evolution and planet formation in protoplanetary disks may proceed differently in diverse metallicity environments. 
Certain aspects of the physical conditions considered by our model could change the way metallicity influences chemistry.

One of the important features is the usage of passive or active disks. For passive disks, the only heating source is stellar radiation and cosmic rays. 
Therefore, passive disks exhibit an increase in temperature with decreasing metallicity. 
In the case of active disks, the hydrodynamical processes provide additional heating  via viscous and compressional heating, which operate predominantly in the disk midplane and depend on the assumed metallicity.
A decrease in metallicity leads to a lower temperature in the midplane (see \fig{fig:disk_init}).

In a recent study, \cite{Guadarrama22} showed that snowlines of chemical species are pushed back as metallicity decreases.
Assuming a passive disk, their model does not take viscous heating into account resulting in an increasing disk temperature throughout the disk with decreasing metallicity.
This assumption might be appropriate for outer disk regions $\gtrsim 10$~AU, where the disk temperature is dominated by stellar irradiation.
Within $R_\mathrm{thick}$, however, the disk midplane temperature decreases with decreasing metallicity (see \sref{sec:disk_init}) and snowlines of species with sublimation temperatures $\gtrsim 50$~K (e.g., H$_2$0 and CO$_2$) are shifted inward.
As a consequence, dust grows and the radial chemical composition of the disk is affected.
We note that our model starts at an already advanced evolutionary disk stage. 
To study the full impact of the aforementioned temperature structures for different metallicities on dust growth and planetesimal formation, the starting point of our model $t_\mathrm{0}$ has to be shifted towards earlier stages.

\subsection{Model limitations}

Although several important aspects of stellar and disk evolution are combined in this study, the results are still preliminary due to certain limitations of our model:

\textit{One-dimensional model:} The main limitation of our disk model is the one-dimensionality. Due to timestep limitations, long-term evolution models that include the innermost disk regions are currently restricted to one dimension \citep[e.g.][]{Steiner21}. 
Unfortunately, several important stellar and disk properties require (at least) an extension in the vertical direction. 
Accretion via funnels, the interaction between stellar magnetic fields and the disk, vertical disk structure, and heating of the disk are intrinsic multi-dimensional processes. 
In the current version of our model, these aspects are included in a simplistic way. 
Nevertheless, the future goal can only be to expand our model by (at least) another dimension.

\textit{Viscous accretion disk:} The transport of mass and angular momentum in our model is described by a simple (constant $\alpha$) viscosity prescription.
$\alpha$ most certainly varies with respect to time and location \citep[e.g.][]{gammie96,King07,Vorobyov20}, and other mechanisms such as disk magneto-centrifugal winds, photo-evaporation and the combination of both also affect the disk evolution \citep[e.g.][]{Konigl11,Roquette21, Kunitomo2020}. These are currently not included but will be added in some future versions.

\textit{Stellar multiplicity and close encounters:} In the current model, only single stars are considered. Binary or even multiple companions can also influence the accretion disk \citep[e.g.][]{Messina17}. Additionally, a fly-by or close encounter with a stellar object can impact the disk evolution \citep[e.g.][]{Vorobyov17x}. These effects, however, are naturally non-axisymmetric events and are currently not feasible with our model.

\textit{Variable dust component:}
The dust-to-gas mass ratio $f_\mathrm{dust}$ in this study is kept constant at the respective stellar metallicity value trough out the simulation and over the disk's radial extent.
While this assumption might be reasonable in the first step, the dust fraction can differ within the disk over time due to a great number of processes, such as coagulation, drift, fragmentation, and vertical settling \citep[e.e.g,][]{Testi2014,Vorobyov2022}.
Our results indicate a direct connection between the amount of dust in the disk and its evolution and a more sophisticated dust evolution model will improve our results.

%
%
%
%
%
%

\section{Conclusion}
\label{sec:conclusion}

In this study, we combine three important aspects of a young star: the stellar evolution with MESA, the disk evolution with TAPIR and the stellar spin evolution based on \cite{Gallet19}. With reference to the influence of stellar metallicity on the evolution of the accretion disk and stellar spin during the star-disk phase, the main conclusions are summarized as follows:

\begin{itemize}
    \item The accretion history and metallicity can influence the evolutionary stage of a star. 
    A low metallicity star is more compact and luminous. 
    The higher the initial accretion rate $\Dot{M}_\mathrm{init}$, the more energy is added to the stellar interior during the early stellar evolution $t<t_\mathrm{0}$ and the larger the stellar radius. On the HRD, the star appears less evolved compared to small values of $\Dot{M}_\mathrm{init}$ (see \fig{fig:star_h_accr} and \fig{fig:star_h_hrd}).
    
    \item Larger luminosities of low metallicity stars affect the radial temperature profile of the accretion disk.
    The disk midplane temperature of the innermost disk region $\lesssim 0.1$~AU as well as the disk beyond $\gtrsim 1$~AU are dominated by stellar irradiation and thus hotter compared to solar metallicity disks.
    Within $R_\mathrm{thick}$, however, the disk midplane is shielded from stellar irradiation and the temperature is dominated by dust opacities. 
    High stellar metallicities results in higher opacity values and the temperatures in the disk midplane are higher compared to lower stellar metallicities (see \fig{fig:disk_init}).
    
    \item The lifetime of an accretion disk is influenced by metallicity. 
    The increased temperatures in the outer disk regions (see \fig{fig:disk_init}) result in a more effective (faster) accretion onto the star and the lifetimes are 0.6-1.4~Myr shorter for low metallicities, within the scope of our model (see \fig{fig:disk_evo}).
    We present an additional explanation for the observed short disk lifetimes in young, low metallicity clusters \citep[][]{Yasui10}. 
    A metallicity-dependent photo-evaporation description \citep[e.g.][]{Nakatani18} arrives at similar results. Their lifetimes, however, are still too long to match the observations. 
    A combination of our star-disk evolution model, coupled with a metallicity-dependent photo-evaporation, could be a key step towards the understanding of disk lifetimes in different metallicity environments.
    
    \item Similar to their older counterparts \citep[e.g.][]{Amard20, Amard20b}, young, low metallicity stars rotate faster compared to solar metallicities (see \fig{fig:spin_evo}).
    We can identify three mechanisms that affect stellar spin evolution. For low metallicities, stellar radii are smaller and, thus, the mechanisms that can remove angular momentum from the star are less effective (see \sref{sec:stellar_spin_model}).
    Additionally, the stellar contraction rate depends on the stellar evolutionary stage.
    The stellar contraction rate is smaller if a star is further evolved. 
    As a consequence, low metallicity stars contract more slowly (see \fig{fig:star_evo_late}) and the spin-up due to contraction is reduced.
    Finally, due to the shorter lifetime of low metallicity disks, stellar spin-up due to contraction, without the interference of a disk, starts at an earlier age. 
    Thus, the difference in rotation period between solar and sub-solar metallicity stars increases further.
    
    \item Including the effects of metallicity, stellar magnetic field strength, and stellar mass in the calculation of the disk lifetime and stellar rotation period during the disk phase, we motivate that our results can be used as input parameters for other recent spin evolution studies \citep[e.g.,][]{Amard19, Amard20, Gossage21}.

\end{itemize}

Our model clearly indicates the potential of the joint hydrodynamic evolution of the star and the disk. 
We motivate the step-by-step inclusion of additional physical aspects (e.g., photo-evaporation and magneto-centrifugal disk winds) and studies covering wider parameter spaces to improve our understanding of these young star-disk systems.

%
%
%
%
%
%


\begin{acknowledgements}

The authors would like to thank the anonymous referee, who provided very detailed comments that helped to improve the quality of the manuscript.
Furthermore, the work of Bill Paxton and his collaborators on the stellar evolution code MESA is gratefully acknowledged.
E.~I.~V. and R.~G. acknowledge support by the Austrian Science Fund (FWF) under research grant P31635-N27. 

\end{acknowledgements}


\bibliographystyle{bibtex/aa}
\bibliography{literature/Tapire}

\begin{appendix}

\FloatBarrier
\section{MESA microphysics}
\label{app:mesaphysics}
The MESA EOS is a blend of the OPAL \citet{Rogers2002}, SCVH
\citet{Saumon1995}, PTEH \citet{Pols1995}, HELM
\citet{Timmes2000}, and PC \citet{Potekhin2010} EOSes.

Radiative opacities are primarily from OPAL \citep{Iglesias1993,
Iglesias1996}, with low-temperature data from \citet{Ferguson2005}
and the high-temperature, Compton-scattering dominated regime by
\citet{Buchler1976}.  Electron conduction opacities are from
\citet{Cassisi2007}.

Nuclear reaction rates are a combination of rates from
NACRE \citep{Angulo1999}, JINA REACLIB \citep{Cyburt2010}, plus
additional tabulated weak reaction rates \citet{Fuller1985, Oda1994,
Langanke2000}. Screening
is included via the prescription of \citet{Chugunov2007}.  Thermal
neutrino loss rates are from \citet{Itoh1996}.

\FloatBarrier
\section{Treatment of rotation in MESA}
\label{app:mesarot}

In MESA, rotation is treated as a modification of the stellar equations due to centrifugal acceleration and non-spherical symmetry, when the star is rotating \citep[see Sec. 6 in][]{paxton2013}.
Angular momentum transport due to rotation (rotational mixing) is implemented as a diffusion term, including five rotationally induced mixing processes: dynamical shear instability, Solberg–Høiland instability, secular shear instability, Eddington–Sweet circulation, and the Goldreich–Schubert–Fricke instability \citep[][]{paxton2013}.

\FloatBarrier
\section{Stellar evolution: $t > t_\mathrm{0}$}

\begin{figure}[ht]
    \centering
         \resizebox{\hsize}{!}{\includegraphics{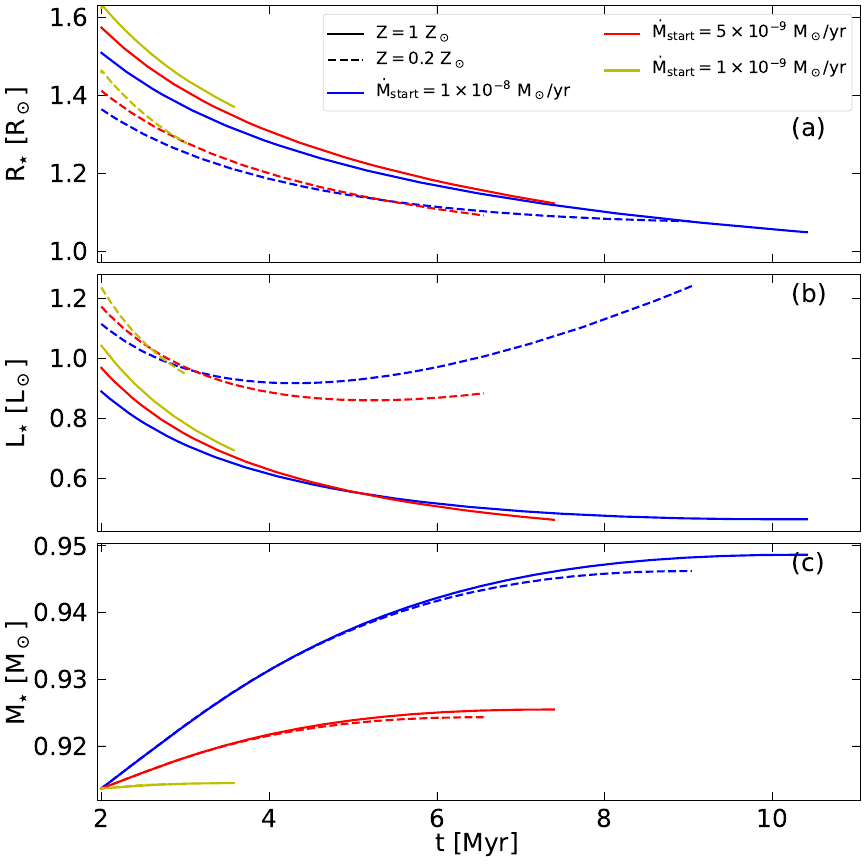}}
    \caption{
    Evolution of the stellar radius (Panel a), intrinsic luminosity (Panel b), and mass (Panel c) for $Z=1~\mathrm{Z_\odot}$ (solid lines) and $Z=0.2~\mathrm{Z_\odot}$ (dashed lines). The yellow, red, and blue lines indicate models with $\Dot{M}_{\rm start} = [1\times10^{-9}, 5\times10^{-9},1\times10^{-8}]~\Msunpyr$, respectively. 
    }
    \label{fig:star_evo_late}
\end{figure}

\fig{fig:star_evo_late} shows the evolution of the stellar radius (Panel a), intrinsic luminosity (Panel b), and mass (Panel c) for $Z=1~\mathrm{Z_\odot}$ (solid lines) and $Z=0.2~\mathrm{Z_\odot}$ (dashed lines). 
The yellow, red, and blue lines indicate models with $\Dot{M}_{\rm start} = [1\times10^{-9}, 5\times10^{-9},1\times10^{-8}]~\Msunpyr$, respectively.

\FloatBarrier
\section{Time evolution of radial disk structure}

\begin{figure*}
    \centering
         \resizebox{\hsize}{!}{\includegraphics{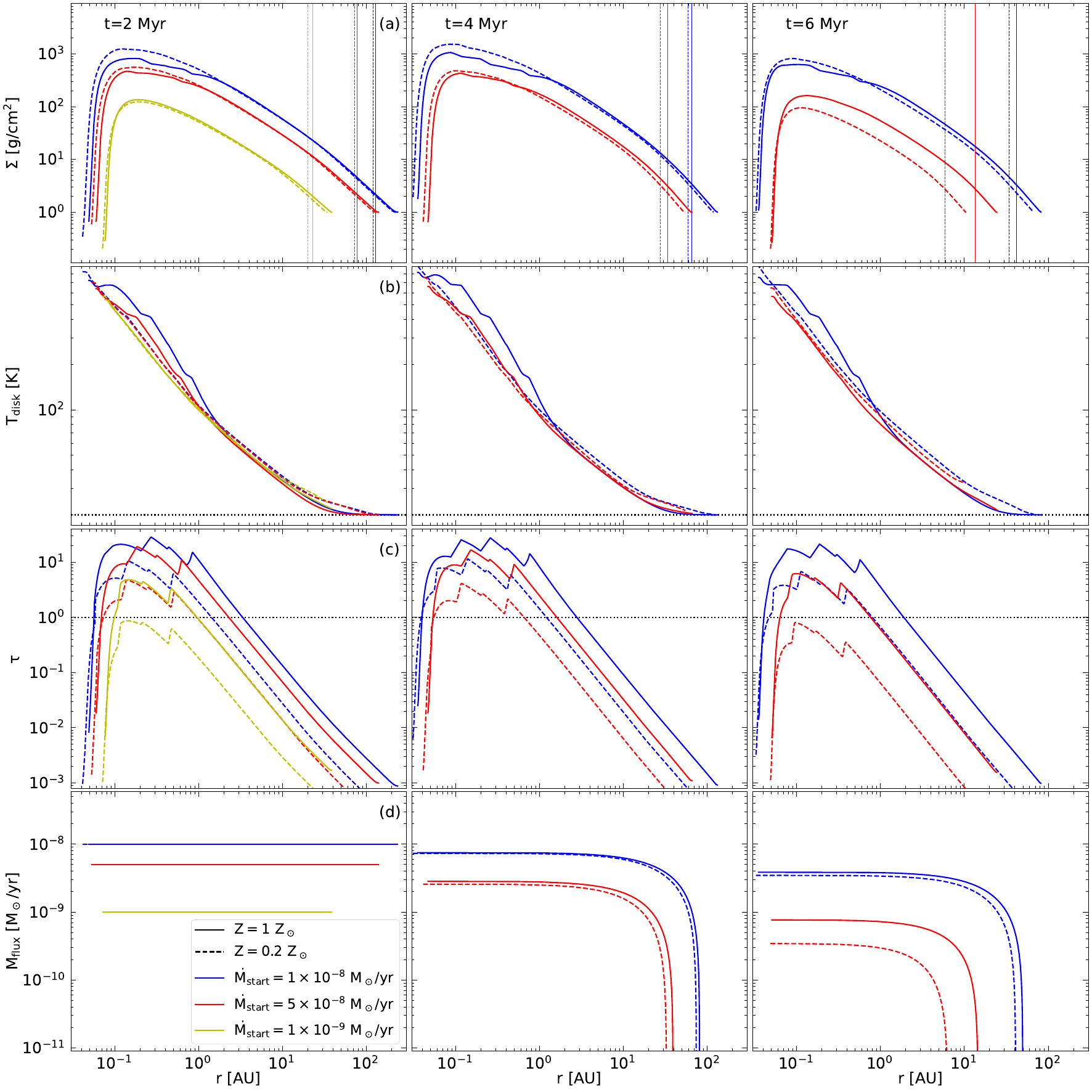}}
    \caption{
    Radial profiles for the model with $\Dot{M}_\mathrm{start} = 1.0\times 10^{-8}~\Msunpyr$ at the starting point of the simulation $t_\mathrm{0} = 2$~Myr (left column), as well as at 5.0~Myr (center column) and 8.0~Myr (right column).
    The solid (dashed) lines correspond to a metallicity of $Z=1~\mathrm{Z_\odot}$ ($Z=0.2~\mathrm{Z_\odot}$), respectively.
    Panels (a) shows the disk surface density $\Sigma$, panels (b) the disk midplane temperature $T_\mathrm{disk}$, panels (c) the optical depth $\tau$ and panels (d) the mass flux $\Dot{M}_\mathrm{flux}$, respectively. 
    The thin vertical lines in panels (a) indicate the respective position of $r_\mathrm{c}$.
    The horizontal lines in panels (b) indicate the ambient temperature $T_\mathrm{amb} = 20$~K.
    The horizontal dashed lines in panels (c) symbolizes $\tau = 1$, dividing optical thin regions ($\tau << 1$) from optical thick regions ($\tau >> 1$).
    }
    \label{fig:disk_rad_comp}
\end{figure*}

The radial profiles for the model with $\Dot{M}_\mathrm{start} = 1.0\times 10^{-8}~\Msunpyr$ at the starting point of the simulation $t_\mathrm{0} = 2$~Myr, as well as at 5.0~Myr and 8.0~Myr are shown in \fig{fig:disk_rad_comp}.
The solid (dashed) lines correspond to a metallicity of $Z=1~\mathrm{Z_\odot}$ ($Z=0.2~\mathrm{Z_\odot}$), respectively.

\FloatBarrier
\section{Age, stellar mass range and metallicity of young clusters}

\begin{table*}[ht]
\centering
\caption{
Age, stellar mass range, and metallicity of young ($\sim 2-5$~Myr) clusters used in \sref{sec:comp_rot}.
}        
\begin{tabular}{c c c c}         
\hline\hline 
Name & Age [Myr] & Stellar mass range [$\mathrm{M_\odot}$] & Metallicity  \\
\hline
NGC~6530 & $\lesssim 5$\tablefootmark{(1)} & $0.2-1.9$\tablefootmark{(2)} & [Fe/H] = 0.11\tablefootmark{(3)} \\
NGC~2362 & $\sim 5$\tablefootmark{(4)} & $0.1-1.2$\tablefootmark{(4)} &  [Fe/H]~$\approx 0$\tablefootmark{(4)} \\
NGC~2264 & $\sim 3$\tablefootmark{(5)} & $0.2-2.5$\tablefootmark{(5)} & [Fe/H] < 0\tablefootmark{(6)} \\
OSFC & $4$\tablefootmark{(7)} & $0.2-1.9$\tablefootmark{(7)} & [Fe/H]~$\approx 0$\tablefootmark{(8)} \\
$\sigma$~Ori & $2-3$\tablefootmark{(9)} & $0.2-0.5$\tablefootmark{(10)} & [Fe/H] = -0.27\tablefootmark{(9)} \\
Taurus & $\sim 3$\tablefootmark{(11)} & $\lesssim 2$\tablefootmark{(11)} & [Fe/H]~$\approx 0$\tablefootmark{(12)} \\

\hline\hline                                            
\end{tabular}

\tablebib{
(1)~\citet{Damiani2019}; (2)~\citet{Henderson12}; (3)~\citet{Tadross2003}; (4)~\citet{Irwin08}; (5)~\citet{Affer13}; (6)~\citet{Cauley2012}, we note that no exact metallicity value is given; (7)~\citet{Serna2021}; (8)~\citet{DOrazi2009}; (9)~\citet{Sherry2008}; (10)~\citet{Cody2010}; (11)~\citet{Rebull2020}; (12)~\citet{DOrazi2011}
}
\label{tab:comp_rot}  
\end{table*}

\FloatBarrier

\end{appendix}

\end{document}